\documentclass[twocolumn]{aastex701}
\usepackage{bm}
\def\lsim{~\rlap{$<$}{\lower 1.0ex\hbox{$\sim$}}}

\def\gsim{~\rlap{$>$}{\lower 1.0ex\hbox{$\sim$}}}

\def\V{\mathcal{V}}
\def\u{{\bf U}} 
\def\cl{C_{\ell}}
\def\N{\mathcal{N}}
\def\thetavec{{\bm{\theta}}}
\def\th{{\bm{\theta}}}

\usepackage{graphicx}	
\usepackage{amsmath}	
\usepackage{psfrag}
\usepackage{bm}
\usepackage{multirow}
\usepackage{ulem}

\begin{document}

\title[Angular Power Spectrum from Galaxy Cluster]{Intensity fluctuations of radio halo in galaxy cluster: Insights from power spectrum estimation}

\author[orcid=0009-0003-3113-2401,gname=Srijita,sname=Pal]{Srijita Pal}
\affiliation{Department of Physics, Indian Institute of Science, Bangalore 560012, India}
\email[show]{srijitapal.phy@gmail.com}  

\author[orcid=0000-0001-9829-7727,gname=Nirupam,sname=Roy]{Nirupam Roy} 
\affiliation{Department of Physics, Indian Institute of Science, Bangalore 560012, India}
\affiliation{Department of Physics, New Mexico Institute of Mining and Technology, Socorro, NM 87801, USA}
\email[show]{nroy@iisc.ac.in}

\author[orcid=0000-0002-0666-2326,gname=Sameer,sname=Salunkhe]{Sameer Salunkhe}
\affiliation{School of Computing, MIT Art, Design and Technology University, Pune, 412201, India}
\affiliation{National Centre for Radio Astrophysics, Tata Institute of Fundamental Research, Pune 411007, India}
\email{sameer24.salunkhe@gmail.com}

\author[orcid=0000-0003-4046-6959,gname=Surajit,sname=Paul]{Surajit Paul}
\affiliation{Manipal Centre for Natural Sciences, Manipal Academy of Higher Education, Manipal, Karnataka 576104, India}
\email{surajit.paul@manipal.edu}

\author[orcid=0009-0005-5899-0486,gname=Tanu,sname=Sharma]{Tanu Sharma} 
\affiliation{Department of Physics, Indian Institute of Science, Bangalore 560012, India}
\email{tanu2023@iisc.ac.in}

\author[orcid=0000-0002-2338-935X,gname=Samir,sname=Choudhuri]{Samir Choudhuri}
\affiliation{Centre for Strings, Gravitation and Cosmology, Department of Physics, Indian Institute of Technology Madras, Chennai 600036, India}
\email{samir.svc@gmail.com}

\begin{abstract}

Non-thermal synchrotron emissions from radio halo allow us to study mechanisms of particle (re)acceleration, magnetic field distribution, merger history, and turbulence in the intra-cluster medium. We propose power spectrum estimation as a novel and complementary method to study galaxy clusters. We use $610\,\,{\rm MHz}$ observations of MACSJ0014.3-302 and MACSJ0152.5-2852 to estimate the angular power spectrum ($\cl$) from the central halo regions. The $\cl$ shows excess emission only for MACSJ0014.3-302. Using simulations, we find that a halo model with power-law fluctuations, in addition to the smooth exponential radial profile, is required to explain the observed $\cl$. We compare the observed power-law with existing models of MHD turbulence. The method may be useful for large data from SKA, finding megahalos in other sources, or detecting faint cluster emissions beyond the visible extent.

\end{abstract}



\section{Introduction}
\label{sec:intro}

Galaxy clusters are the largest gravitationally bound structures in the Universe, consisting of hundreds of galaxies, and hot, diffuse, magnetized plasma (called the intra-cluster medium; ICM) at a temperature of $10^7 - 10^8\,\,{\rm K}$ \citep{Krav12,Lov24}. Galaxy clusters exhibit diffuse X-ray emissions by thermal Bremsstrahlung from the hot ICM plasma that fills the entire cluster volumes, and faint radio emissions from synchrotron radiations that can extend across 100s of kpc to a few Mpc scale. They present a unique environment for studying particle acceleration, particularly cosmic-ray electrons, thermal plasmas, shocks, and turbulence, which are crucial in energizing these electrons to diffuse out to a larger distance from the cluster centre \citep{Sar02,Paul11,Brun14,Wee19}. Radio observations provide an alternative method to detect and study the galaxy clusters along with the X-ray, SZ, or optical surveys \citep{Wee19}. The non-thermal synchrotron emissions also indicate towards the presence of cosmic-ray electrons and large-scale ($\sim{\rm Mpc}$) magnetic fields in the ICM unassociated with the individual radio galaxies within the clusters. Cosmic-ray electrons, in the presence of magnetic fields with a strength of $0.1-10\,\,\mu{\rm G}$ \citep{Wee19}, give rise to non-thermal synchrotron radiation; how this radiation is sustained to such huge distances ($\sim {\rm Mpc}$) in the galaxy clusters, however, is still unclear, and requires a continuous injection or acceleration of electrons across the entire galaxy cluster \cite{Jaffe77,Bot22,Wee17} through processes like adiabatic compression, Fermi-I and Fermi-II acceleration. Observing this radio emission helps reveal the underlying physical processes behind the cosmic-ray electron propagation and magnetic fields in the ICM.

The various cluster-wide diffuse radio emission that have been observed so far are classified into radio halos and mini-halos, radio relics, and fossil plasma remnants of active galactic nuclei (\cite{Wee19,Paul23}). Galaxy clusters form through highly energetic, complex merger events that introduce shocks and turbulence in the ICM. These likely amplify the ICM magnetic fields, by shock compression or turbulent dynamo, that control particle acceleration, turbulence and the non-thermal synchrotron emission in the ICM. The same shock and turbulence, generated through merger events, also likely drive the underlying (re-)acceleration mechanisms of the cosmic-ray electrons, that produce these radio structures in the galaxy clusters \citep{Brun14}. Thus, the diffuse radio emission is quite useful to study the ICM turbulence as well as shocks, providing a unique probe to the merger histories and the dynamical state (relaxed or merging) of the galaxy clusters. There are no direct methods currently to observe ICM turbulence; synchrotron emission is the only way for an indirect, but reasonable, detection.

Radio halos are diffuse and low surface brightness ($\sim1\mu {\rm Jy}/{\rm arcsec}^2$ spanning $\sim$ 1 Mpc at 1.4 GHz) emission in the central part of galaxy clusters. Further, they have a steep spectrum with $\alpha \geq 1 $ between 300 MHz and GHz. Studies \citep{Br09,Ro11,Cu18} suggest that the presence of radio halos are strongly related to the evolutionary and merging state, being more common in merging clusters and lacking in relaxed systems. Their origin is thought to be linked with production and/or (re-)acceleration of cosmic-ray electrons as well as magnetic field amplification through shocks and turbulence in the ICM induced by the merger events. Thus, radio halos provide a huge opportunity to study turbulence.

Despite its importance, astrophysical turbulence remains poorly understood, and study of the galaxy cluster halo emission and its connection to the nature of the underlying turbulence can be an important step forward. Focusing on the inertial range of the energy cascade, existing turbulence models, e.g. Kolmogorov model for incompressible, isotropic, and non-magnetic turbulence \cite{Ko41}, Iroshnikov-Kraichnan (IK) model in presence of strong magnetic field \citep{Ir64,Kr65}, or Goldreich-Sridhar (GS) model of anisotropic strong magnetohydrodynamic (MHD) turbulence \citep{GS95}, all predict a power-law scaling of the energy spectrum but with different power-law index. \cite{Ber06} demonstrated that polarization intermittency in strong MHD turbulence results in a shallower power-law scaling with increased anisotropy than that predicted in the Goldreich-Sridhar model. The multi-dimensional form of Burger's equation has been used to introduce topological shocks in forced Burgers turbulence, which also predicts power-law scaling \citep{Bec07}. In the computational front \citep{Cho02,Cho02Ph,Cho02ApJ,Ber06}, understanding on compressible MHD turbulence has progressed through detailed numerical simulations of specific situations, demonstrating importance of anisotropy and intermittency (also see \cite{Min15} and \cite{Ber16} and references therein for the challenges and current status specifically regarding simulations of MHD turbulence in the ICM). Thus, careful observation, modelling and comparison of the observed power spectrum with the predictions from analytical or numerical work can be used to constrain the model parameters and help understanding the nature of astrophysical turbulence.

The detection of radio halos is primarily limited by instrument sensitivity \citep{Gio09,Paul19}. Moreover, recent detections of the megahalos \citep{Cuc22,Salunke25} suggest that the magnetic field and cosmic-ray electrons occupy a much larger volume than previously observed. Megahalos are huge (30 times larger in volume than the typical radio halo), low surface brightness and steep spectrum radio structures that can extend far out to $2-3\,\,{\rm Mpc}$. So far, five detections of megahalos have been reported, which further complicates the scenario in explaining the origin of such extended diffuse radio emission in galaxy clusters \citep{Cuc22,Salunke25}. These authors have concluded that the presence of megahalos may not be very unique and it may be reasonably detected  in many other clusters. Possibly, they currently remain undetected because of their ultra-steep nature and the lack of instrument sensitivity.

Apart from sensitivity, current limitations in galaxy cluster studies include data analysis challenges, e.g. deconvolution error, image artifacts in presence of bright sources etc. There are some recent progress with deeper observations, polarimetric as well as multi-frequency studies \citep{Wee19}. Upcoming instruments like the Square Kilometer Array (SKA; \cite{Br15}) will improve it further, but will also pose more challenges in terms of data volume and analysis strategies. In this work, we consider the angular power spectrum (APS, $\cl$), which quantifies the statistics of the sky signal as a function of the angular multipole $\ell$, to study the radio halos of galaxy clusters. This second order statistics is a powerful tool widely used across disciplines including astrophysics and cosmology (e.g. CMB, 21-cm cosmology; \citep{laporta08,samir17a}).

The Tapered Gridded Estimator \citep[TGE;][]{samir14,samir16,samir17} is a novel power spectrum estimator that has been developed primarily in the context of $21$-cm cosmology \citep{BNS}, and later applied for various other studies. The salient features of the TGE are as follows, (1) it estimates the power spectrum by convolving and gridding the visibilities in the uv-domain (Fourier domain of the image plane), consequently tapering the antenna response of the telescope in the image domain at wide angular distances outside the main lobe of the telescope primary beam (PB) pattern. The TGE thus suppresses the contributions from the wide-field foreground contamination and mitigates its effects from the estimated power spectrum; (2) the TGE internally estimates and subtracts out the positive noise-bias from the visibilities to provide an unbiased estimate of the measured quantity; (3) it also reduces the computational load due to gridding the visibilities to make the estimation, an important factor for future telescopes like SKA which are expected to produce large amount of data.

So far, several studies have used 2D TGE to measure the angular power spectrum $C_{\ell}$ of the Diffuse Galactic Synchrotron Emission \citep[DGSE;][]{shaver99} at $150 \, {\rm MHz}$  using GMRT \citep[Giant Metrewave Radio Telescope;][]{swarup91} TGSS observations \citep{samir17a,samir20}. \citet{Cha1,Cha2} and \citet{M20} have also used the 2D TGE to measure $C_{\ell}$ of the DGSE at $325\,\,{\rm MHz}$. The 2D TGE have also been used to measure $C_{\ell}$ of the fluctuations in the synchrotron emission from the Kepler supernova remnant and study the magnetohydrodynamic turbulence in the supernova remnants \citep{Preetha19, Preetha21}. \citet{ITGE} have developed an Image-based Tapered Gridded Estimator (ITGE) which was used to measure $C_{\ell}$  of the HI 21-cm emission from the ISM in different parts of of an  external galaxy. These studies clearly establish the 2D TGE as an efficient and reliable estimator of the angular power spectrum $C_{\ell}$; they also demonstrate its ability to suppress the foreground contribution far away from the phase center. Further, the multi-frequency power spectrum $C_{\ell}(\nu_{a},\nu_{b})$ \citep[MAPS;][]{KD07} based TGE \citep{Bh18} has been extensively studied using GMRT and upgraded GMRT data in the context of Epoch of Reionization (EoR) and post-EoR 21-cm power spectrum estimation. These studies \citep{Pal20,Pal22} clearly demonstrated all the capabilities of the TGE mentioned above, as well as provided  $2\sigma$ constraints on $\Delta^{2}(k)$ the amplitude of the mean squared HI 21-cm brightness temperature fluctuations at EoR ($z=8.28$) and post-EoR ($z=2.28$), where $z$ denotes the redshift of observation.

We propose that the power spectrum of individual galaxy clusters can reveal structures of the intensity fluctuations present at Mpc scales, and constrain the physical processes including turbulence that generate these fluctuations. Here, for the first time, we estimate the $\cl$ for the specific intensity fluctuations from the radio halos. We have used a fast and novel method where the two-point correlations of the visibilities, which are the directly measured quantity in radio interferometric observations, yield an estimate of $\cl$ of the specific intensity fluctuations \citep{BNS}. This mitigates many of the complications of data analysis and challenges with data volume mentioned above. In this paper, we first present and establish the methods developed to perform the $\cl$ analysis for the galaxy clusters in detail. So far, 2D TGE has been primarily used to characterize the diffuse emission that essentially fills the entire field-of-view (FoV) of the observing telescope; in this work, we first adapt the 2D TGE for sources restricted within a small part of the PB, which is usually the case for radio halos (relics too). We study the effect of the emission restricted within a small part of the PB on TGE, and find that an additional amplitude correction enables us to adapt the 2D TGE for such sources, which we formulate and verify here. Additionally, we have presented a comprehensive method to estimate the DGSE from any FoV. Note that, the tapering feature in the TGE suppresses the residual compact (point) source contributions that would otherwise dominate the estimated angular power spectrum. This also allows us to probe a wider range of angular multipoles in our studies. Finally, even though the entire methodology presented in the paper is used to study $\cl$ for galaxy clusters, this can be used for any similar cases where the emissions are restricted within a small part of the PB. We plan to use this in our future studies involving dark matter detection from dwarf spheroidal galaxies. Finally, we have used the estimated $\cl$ to constrain the turbulence in the ICM for Abell 2744.

This paper is arranged as follows. We describe the GMRT observations and the initial processing of the data in Section \ref{sec:obs}. The methodology is described in detail in Section \ref{sec3}, and is validated using simulations in Section \ref{sec:simulation}. We present the results in Section \ref{sec:sec3}, and the discussions in Section \ref{sec:sec4}. We summarize and conclude in Section \ref{sec:sec5}.

\section{Observation and Data Analysis}
\label{sec:obs}

\begin{table}
\caption{Observation summary. Notes: $a$: \cite{Ebe10}, $b$: \cite{Horesh10}}
\label{t_1}
\resizebox{\linewidth}{!}{
\begin{tabular}{lcc}
\hline
\hline
Central Frequency $(\nu_c)$ & & $612.33$ MHz  \\
\hline
Channel width $(\Delta\nu_c)$ & & $1497.40$ kHz \\
\hline
Bandwidth $(B_{bw})$ & & $32.94$ MHz \\
\hline
Target field  $(\alpha,\delta)_{2000}$ & Abell 2744 & ($00^h14^m18.89^s$,$-30^{\circ}23^{'}22.46^{''}$) \\
 & MACSJ0152 & ($01^h52^m34.49^s$,$-28^{\circ}53^{'}36.05^{''}$) \\
\hline
Galactic coordinates $(l,b)$ & Abell 2744 & ($8.90^{\circ}, -81.24^{\circ}$) \\
 & MACSJ0152 & ($223.95^{\circ}, -76.40^{\circ}$) \\
\hline
Redshift $(z)$ & Abell 2744 & $0.308^{a}$ \\
 & MACSJ0152 & $0.413^{b}$ \\
\hline
GMRT Observation date & Abell 2744 & 13-AUG-2011 \\
 & MACSJ0152 & 05-JUL-2011 \\
\hline
On-source observation time & Abell 2744 & 05:31 hour \\
 & MACSJ0152 & 04:50 hour \\
\hline
\hline
\end{tabular}
}
\end{table}

For this work, we have considered Abell 2744 and MACSJ0152 from the MACS catalogue \citep{Ebe10,Paul19}, which have also been studied in our previous work \citep{Paul19}. We have used GMRT observations of the two galaxy clusters at $610\,\,{\rm MHz}$. We have chosen the clusters considering that (1.) the high dynamical activities observed in them allow us to study the ICM at different phases of mergers \citep{Zitrin11}; (2.) the GMRT observations considered here show significant emission above the observed noise level in the radio map for Abell 2744, whereas, discernibly visible emission cannot be seen for MACSJ0152 \citep{Paul19}, which enables us to test our power spectrum methodology for the galaxy clusters that we have proposed in the paper.

\subsection{GMRT observations and initial data processing}

The galaxy clusters were observed using the GMRT radio interferometer located in the western parts of India. The observations were carried out during July-August 2011 (Project Code: 20\_062), in dual-band mode at 235 MHz and 610 MHz, where Stokes ``RR" was used for 610 MHz and ``LL" for 235 MHz; we have only considered 610 MHz data for this work. The relevant observation and data parameters are summarized in Table \ref{t_1}; for more details, the reader is referred to \citealt{Paul19}. The visibilities were recorded over the 32 MHz bandwidth for both sources being investigated.

The raw visibilities were processed using the fully automated {\scriptsize SPAM} pipeline \citep{Intema17} that included several rounds of flagging, (self)calibration, and wide-field imaging \citep{WhiteBook}. Source 3C48 was used for the initial flux and bandpass calibrations for both target sources. The calibration part also included direction-dependent calibrations, which corrected for ionospheric delays and variations in the antenna beam pattern.

\subsection{Imaging and Self-calibration}

\begin{table*}[ht!]
\centering
\caption{Imaging summary. Column `I' corresponds to before self-calibration, Column `II' corresponds to after self-calibration and Column `III' corresponds to after compact source subtraction.}
\label{t_Im}
\begin{tabular}{lcccc}
\hline
\hline
 & Source & I & II & III \\
\hline
\hline
Image size & & $2048\times2048$ & $2048\times2048$ & $2048\times2048$ \\
\hline
Pixel size & & $3\times3\,\,{\rm arcsec}^{2}$ & $3\times3\,\,{\rm arcsec}^{2}$ & $3\times3\,\,{\rm arcsec}^{2}$ \\
\hline
Robust & & 2.0 & 2.0 & 2.0 \\
\hline
uvrange & & 0-30 kL & 0-30 kL & 0-30 kL \\
\hline
UV taper & & [`15 kL', `15 kL', `0deg']  & [`15 kL', `15 kL', `0deg']  & [`15 kL', `15 kL', `0deg'] \\
\hline
Off-source r.m.s & Abell 2744 & $66$ & $50$ & $50$ \\
$\mu{\rm Jy/beam}$ & MACSJ0152 & $45$ & $41$ & $41$ \\
\hline
Synthesized beam & Abell 2744 & $26\times16\,\,{\rm arcsec}^{2}$, PA $=13^{\circ}$ & $26\times16\,\,{\rm arcsec}^{2}$, PA $=12^{\circ}$ & $26\times16\,\,{\rm arcsec}^{2}$, PA $=12^{\circ}$ \\
& MACSJ0152 & $27\times13\,\,{\rm arcsec}^{2}$, PA $=14^{\circ}$ & $27\times13\,\,{\rm arcsec}^{2}$, PA $=14^{\circ}$ & $27\times13\,\,{\rm arcsec}^{2}$, PA $=14^{\circ}$ \\
\hline
\hline
\end{tabular}
\end{table*}

\begin{figure*}
\begin{center}
\includegraphics[scale=0.45]{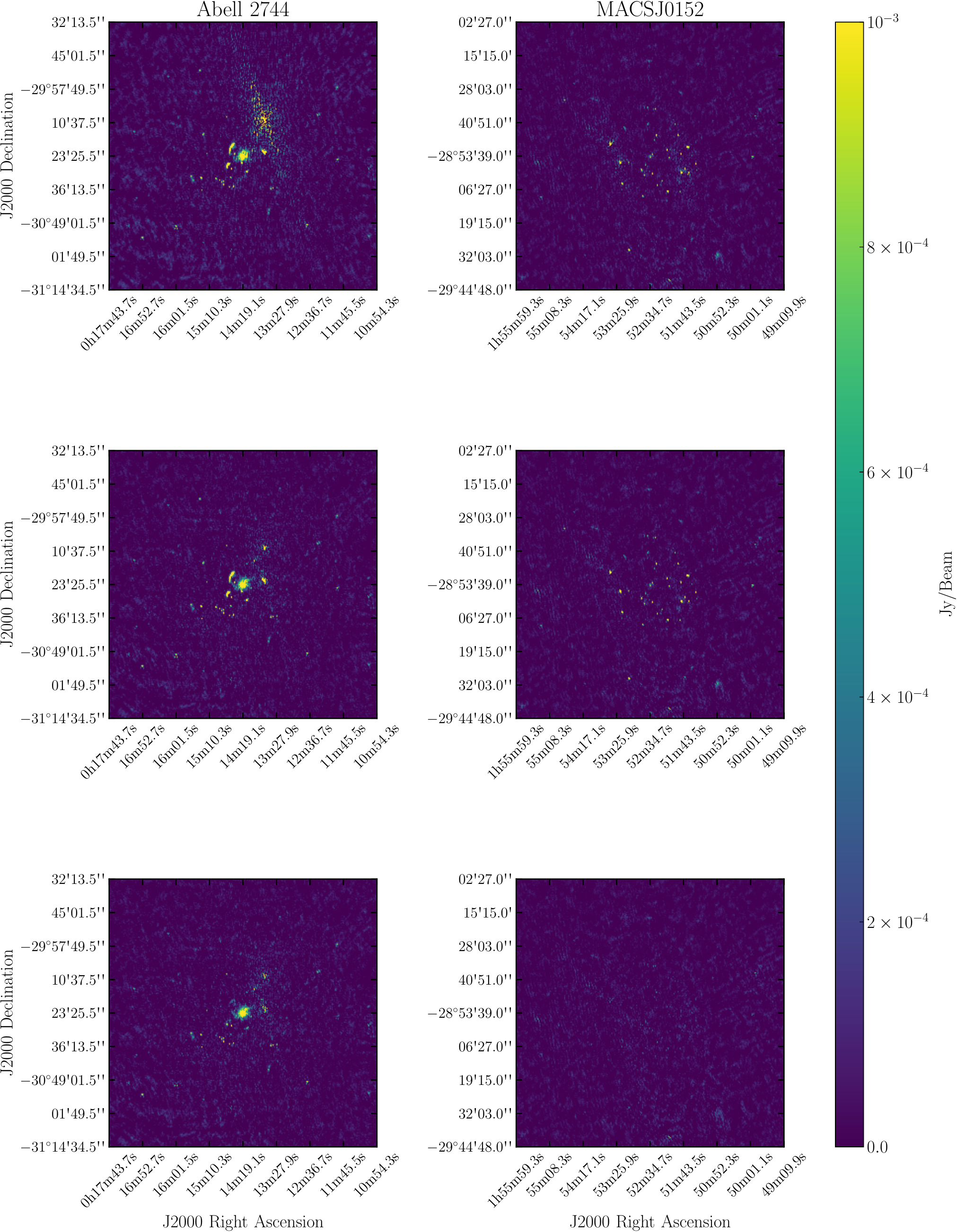}
\caption{Total intensity map of Abell 2744 (left column) and MACSJ0152 (right column) fields at $\sim 610$ MHz, obtained using {\scriptsize CASA} task {\scriptsize tclean}, after - (uppermost row) initial processing of the data with {\scriptsize SPAM}; (middle row) additional rounds of manual self-calibration using {\scriptsize CASA}; (lowermost row) compact source subtraction from the data.}
\label{fig:fig1}
\end{center}
\end{figure*}

We have used {\scriptsize CASA} task {\scriptsize tclean} to make an initial image of the fields. Column `I' of Table \ref{t_Im} gives a summary of the imaging parameters, and we have shown the corresponding images in the uppermost panels of Figure \ref{fig:fig1}. The GMRT FoV at $610$ MHz is relatively small ($\theta_{\rm FWHM}\sim43^{'}$); a standard gridding (gridder=`standard' in {\scriptsize CASA} task {\scriptsize tclean}) sufficed for us, given that we mainly analyzed the data in the visibility plane and that our estimator can also mitigate wide field effects in the data. We have used Briggs weighting with robust parameter 2, which nearly corresponds to a natural weighting and provides higher sensitivity in the image. We have also used a circular Gaussian uv taper of $15\,{\rm kL}$ to obtain smoother PSF; this helps us reduce sidelobes, suppress artifacts and bring out the diffuse emissions better. We have combined the frequency channels using the multi-frequency synthesis to obtain an image at a single channel. We have made the image ($\sim 102^{'}\times102^{'}$) slightly larger than the FoV to also display the bright sources outside the main lobe of the FoV. 

The image (uppermost row of Figure \ref{fig:fig1}) shows artifacts at wide angular scales, around the two brightest sources in the first field centering Abell 2744, despite the calibrations performed in {\scriptsize SPAM}. This showed that the automated calibration performed in {\scriptsize SPAM} is not optimal. To mitigate this, we have performed additional manual self-calibration in order to improve the complex gains and better model the compact sources. The accurate modelling and removal of the compact sources is important, considering that our target here is to study the galaxy cluster emissions, which are not the brightest sources in the respective fields. For Abell 2744, we have performed four rounds of phase-only self-calibration \citep{WhiteBook} using the brightest sources present in the field with a solution time interval of 8,5,2 and 1 min for the successive self-calibration loops. With each loop, we see improvements in the gain solutions, as well as in the SNR ($\approx 1862,\,4425,\,6918,\,7131$ respectively considering off-source noise from a random region in the sky). For MACSJ0152, only one round of phase-only self-calibration with a solution time interval of 8 min was adequate; after this the SNR saturates and no further improvements are observed. The final continuum images after these manual self-calibration steps are shown in the middle panels of Figure \ref{fig:fig1}; the relevant parameters for imaging after these additional self-calibrations are tabulated in the Column `II' of Table \ref{t_Im}. The resulting improvements are prominent around the bright sources, and we find that the remaining artifacts due to residual phase errors have no impact on the subsequent analyses. The off-source r.m.s. and the size of the synthesized beam are given in Table \ref{t_Im} (Column `II'). Note that, the off-source r.m.s. has some variation across the sky; we have considered seven representative, off-source and noise dominated regions in the images to estimate the r.m.s. in Table \ref{t_Im} for both sources. In any case, we see significant improvement than before, just as seen from Column `II' of Table \ref{t_Im} relative to Column `I'. We have applied the final gain table to all frequency channels for both the sources. 

\subsection{Point source modelling and Foreground subtraction}
\label{sec:pssub}

The middle row of Figure \ref{fig:fig1} shows the self-calibrated images for the two fields. We notice that the fields are dominated by the discreet and compact objects within the $102^{'}\times102^{'}$ FoV of the images at 610 MHz. The radio halo, as well as the relics, shows significant emissions above the noise level for Abell 2744, whereas no visible radio emission is observed for MACSJ0152 at 610 MHz. Considering Abell 2744, the brightest sources within the FoV other than the central halo are identified as radio relics of the galaxy cluster itself, other individual radio galaxies, candidate active galactic nuclei (AGN) etc. For MACSJ0152 we see that the discrete point sources are quasars, AGN, radio galaxies etc. We term all the compact sources and all-sky diffuse emissions brighter than the diffuse emission from the radio halos being studied, as foregrounds for both fields. The foregrounds dominate the angular power spectra over the halo emissions from the clusters, if not modelled and removed properly from the visibility data. We have used the {\scriptsize CASA} task {\scriptsize tclean} to obtain a model of these foregrounds, which we have subsequently subtracted from the visibilities using {\scriptsize uvsub} in {\scriptsize CASA}. For Abell 2744, we have used {\scriptsize tclean} interactively and iteratively to manually identify all the sources within the FoV which are brighter than the radio halo situated at the centre of the FoV, including the radio relics. Here, our goal was to identify the brighter sources in a very conservative manner so that diffuse emissions from the halo are not removed in the process. We could achieve it by carefully selecting and taking out only the brightest sources one-by-one using a few rounds of {\scriptsize tclean} interactively in CASA. We then removed the corresponding visibilities from the total data. The residual data ideally contain the radio halo, and all compact and diffuse emissions fainter than the radio halo. For MACSJ0152, we have modeled and removed all bright sources above a threshold flux level of $\sim 0.5$ mJy/Beam. We expect the residual data to contain the diffuse emissions, noise and the unsubtracted point sources that lie below this. For both clusters, we have been cautious to avoid over-subtraction which can potentially lead to signal loss in the data; we have taken a conservative approach for the foreground subtraction, so as not to mistakenly also remove any diffuse emission from the radio halos being investigated.

The lowermost panels of Figure \ref{fig:fig1} show the fields after the foregrounds have been subtracted from the data; the relevant imaging parameters are tabulated in Column `III' of Table \ref{t_Im}. We see that for both fields we have successfully subtracted out most of the genuine foregrounds otherwise observed in the self-calibrated images. For Abell 2744, the residual emissions observed around the brightest sources after the foregorund removal have little effect on our results. We find that after foreground subtraction sources above the flux cut-off threshold $S_{c} = 3.5\,\,{\rm and}\,\,1\,\,{\rm mJy/beam}$ have been removed from the fields for Abell 2744 and MACSJ0152 respectively. The peak/noise in the final images are $60\,\,{\rm and}\,\, 24$ respectively corresponding to the r.m.s. noise $50\,\,{\rm and}\,\,41\,\,\mu{\rm Jy/beam}$ respectively for Abell 2744 and MACSJ0152. The self-calibrated visibilities before and after the foreground subtractions within a baseline range of $\lvert {\bf U}_{i} \rvert \leq 3000 \lambda$ have been used in the subsequent analysis.

\section{Methodology}
\label{sec3}

\subsection{The Tapered Gridded Estimator for the Angular Power Spectrum}

The fluctuations in the brightness temperature distribution across the sky $\delta T_{\rm b} (\hat{\bm{n}},\,\nu)$ at a frequency $\nu$ can be decomposed in terms of spherical harmonics $Y_{\ell}^{\rm m}(\hat{\bm{n}})$ as
\begin{equation}
\delta T_{\rm b} (\hat{\bm{n}},\,\nu)=\sum_{\ell,m} a_{\ell {\rm m}} (\nu) \,
Y_{\ell}^{\rm m}(\hat{\bm{n}}).
\label{eq:alm}
\end{equation}
The angular power spectrum is then defined as
\begin{equation}
C_{\ell} = \big\langle a_{\ell {\rm m}} (\nu)\, a^*_{\ell
  {\rm m}} (\nu) \big\rangle\,
\label{eq:cl}
\end{equation}
where the angular brackets $\langle  ... \rangle$ denote an ensemble average over different statistically independent realizations of the Gaussian random field  $\delta T_{\rm b} (\hat{\bm{n}},\,\nu)$. If we assume that $\delta T_{\rm b} (\hat{\bm{n}},\,\nu)$ are generated as a result of a Gaussian random process that is statistically homogeneous and isotropic in the sky, this second order statistic completely quantifies the statistical properties of the sky signal.

We use 2D TGE to measure $\cl$ from the residual visibilities after foreground subtraction. This estimator has been derived and discussed extensively in previous works and here we have presented a working summary; readers may refer to \citealt{samir14,samir16} for details. The TGE uses the complex visibilities $\V_i$, measured at an $i$-th baseline $\u_i$ and at a single frequency $\nu$, to estimate the convolved gridded visibilities $\V_{cg}$ by convolving $\V_{i}$ with a function $\tilde{w}(\u)$. We consider a rectangular grid in the $uv$ plane and evaluate $\V_{cg}$ at baselines $\u_g$ corresponding to the different grid points using
\begin{equation}
	\V_{cg} = \sum_{i}\tilde{w}(\u_g-\u_i) \, \V_i \,.
	\label{chap2:eq:a1}
\end{equation}
The function $\tilde{w}(\u)$ is the Fourier transform of a window function ${\cal W}(\thetavec)$, which here has the form
\begin{equation}
	{\cal W}(\thetavec)=e^{-\theta^2/\theta_w^2}\,.
	\label{eq:win}
\end{equation}
The parameter $\theta_w$ controls the width of the Gaussian and is used to restrict ${\cal W}(\thetavec)$ to within the first null of the primary beam pattern of the telescope \citep{samir14}. The convolution in $\V_{cg}$ (eq. \ref{chap2:eq:a1}) tapers the sky response at wide angles away from the phase centre and suppresses foreground contributions from the outer regions and side-lobes of the telescope's primary beam pattern \citep{samir14,samir16}, and the gridding in $\V_{cg}$ reduces computation. We have further combined the visibilities in different frequency channels to a single grid, and also scaled each baseline appropriately in frequency for gridding.

We have used $\theta_w = f \theta_0$, where `$f$' is the tapering parameter and $\theta_{0}=0.6\times\theta_{\rm FWHM}$, $\theta_{\rm FWHM}$ being the full width at half maxima of the primary beam of the telescope at the observing frequency. Note that, $f < 1$ produces large tapering, whereas a large value $f > 1$ implies very little tapering and we consider $f=10.0$ to be equivalent to no tapering being applied. For reference, considering GMRT at $610 \, {\rm MHz}$ we have $\theta_{\rm FWHM}\sim43^{'}$, whereas $\theta_w\approx15^{'}$ at $f=0.6$ which is the highest tapering we have used for this work.

The 2D TGE estimator is defined as 
\begin{equation}
	{\hat E}_g= M_g^{-1} \, \left( \mid \V_{cg} \mid^2 -\sum_i \mid
	\tilde{w}(\u_g-\u_i) \mid^2  \mid \V_i \mid^2 \right) \,,
	\label{eq:a2}
\end{equation}
with $\langle {\hat E}_g \rangle = \cl{_g}$ where $\ell_g=2 \pi U_g$, and $\langle ... \rangle$ denotes an ensemble average over multiple realizations of the sky brightness temperature fluctuations. The second term in the brackets $(...)$ in eq.~(\ref{eq:a2}) subtracts out the noise bias contribution and the estimator yields unbiased estimates of $\cl$. The normalization factor $M_g$ is estimated through simulation (section and reference). Finally, we average over isotropic annular bins of equal logarithmic width in the $uv$ plane to obtain the binned average $\cl$ corresponding to the effective angular multipoles $\ell$.

\subsection{Normalization due to finite size of the Galaxy Clusters}
\label{sec3b}

In this work, we have further adapted the TGE to estimate the $\cl$ from sky emissions restricted within a small part of the PB, which essentially amounts to introducing correct normalization which accounts for the finite extent of the target sources. We present the formalism in this subsection. Here, we study the effect of such emission from a finite region on the TGE, and formulate the additional amplitude correction that enables us to adapt the 2D TGE for such sources.

We represent the brightness temperature fluctuations for the source restricted within a small part of the PB as $[{\cal F}(\bm{\theta})\,\delta T_{b}(\bm{\theta})]$, where, $\delta T_{b}(\bm{\theta})$ is an all-sky diffuse background field representing the underlying distribution of the brightness temperature fluctuations from the galaxy cluster, and we incorporate the finite size of the galaxy cluster by modulating the underlying $\delta T_{b}(\bm{\theta})$ with a suitable window function represented by ${\cal F}(\bm{\theta})$. Here we work under flat-sky approximation, which is reasonable considering that the GMRT FoV is small ($\sim 43^{'}$) at $610\,\,{\rm MHz}$, and $\bm{\theta}$ represents the two dimensional position vector on the sky plane perpendicular to the line-of-sight of the observation denoting the angular distance from the phase centre. We have chosen ${\cal F}(\theta)=e^{-\theta^{2}/\theta_{1}^{2}}$, where $\theta_{1}=2\,r_{h}$ in radian, $r_{h}$ being the half-light radius of the source, and $\theta=\mid\bm{\theta}\mid$.

At a particular frequency, the observed field in the Fourier domain due to the source after TGE is then given by,
\begin{equation}
\Delta T_{b}^{f}(\u)=\int_{-\infty}^{\infty}\,{\cal W}(\bm{\theta})\,{\cal A}(\bm{\theta})\, [{\cal F}(\bm{\theta})\,\delta T_{b}(\bm{\theta})]\,\,e^{+2\pi i \u\cdot\bm{\theta}} d^{2}{\theta}.
\label{eq:c1}
\end{equation}
Considering eq.~(\ref{eq:c1}), we have explicitly introduced the PB pattern ${\cal A}(\bm{\theta})$ and the tapering window function in the TGE ${\cal W}(\bm{\theta})$. We have already defined ${\cal W}(\theta)=e^{-\theta^{2}/[f\theta_{0}]^{2}}$ and discussed it above (see eq.~\ref{eq:win}). Additionally, for a circular antenna, ${\cal A}(\theta)$ is well-modelled by a Gaussian function of the form ${\cal A}(\theta)=e^{-\theta^{2}/\theta_{0}^{2}}$ within the main lobe of the FoV \citep{samir14}.

It is convenient to represent eq.~(\ref{eq:c1}) in terms of convolution as,
\begin{equation}
\Delta T_{b}^{f}(\u)=\int_{-\infty}^{\infty}\,\tilde{w}^{\rm eff}(\u-\u^{'})\, \Delta T_{b}(\u^{'})\,\,d^{2}{\rm U}^{'}\,.
\label{eq:c2}
\end{equation}
Here, $\Delta T_{b}({\u})=F[\delta T_{b}(\bm{\theta})]$, and the effective window for the halo of the cluster is defined as $\tilde{w}^{\rm eff}(\u)=F[{\cal F}(\bm{\theta})\,{\cal W}(\bm{\theta})\,{\cal A}(\bm{\theta})]$, where `$F$' represents a Fourier transform. The product of the three terms within the square brackets $[...]$, can then be denoted by an effective Gaussian of the form, ${\cal F}(\bm{\theta})\,{\cal W}(\bm{\theta})\,{\cal A}(\bm{\theta})=e^{-\theta^{2}/\theta_{\rm eff}^{2}}$, with $\frac{1}{\theta_{\rm eff}^{2}}=\frac{1}{\theta_{1}^{2}}+\frac{1}{[f\theta_{0}]^{2}}+\frac{1}{\theta_{0}^{2}}$, and $\tilde{w}^{\rm eff}({\rm U})=\frac{1}{\pi\, {\rm U}^{2}_{\rm eff}}\,e^{-{\rm U}^{2}/{\rm U}^{2}_{\rm eff}}$ where ${\rm U}_{\rm eff}=\frac{1}{\pi\,\theta_{\rm eff}}$.

The measured angular power spectrum of the brightness temperature fluctuations for the source restricted within a small part of the PB, $C_{\ell=2\pi{\vert\u\vert}}^{f} = \langle \Delta T_{b}^{f}({\u})\Delta T_{b}^{*f}({\u}^{'})\rangle$, is then given by,
\begin{equation}
\begin{split}
\cl^{f} = \int_{-\infty}^{\infty}\int_{-\infty}^{\infty}\,\tilde{w}^{\rm eff}&(\u-\u^{''})\tilde{w}^{*{\rm eff}}(\u-\u^{'''})\\
 &\langle\,\Delta T_{b}({\u}^{''})\Delta T_{b}^{*}({\u}^{'''})\,\rangle\,\,d^{2}{\rm U}^{''}d^{2}{\rm U}^{'''}
\label{eq:c3}
\end{split}
\end{equation}
The actual underlying angular brightness temperature power spectrum $\cl$ is defined as,
\begin{equation}
\langle\,\Delta T_{b}(\u)\Delta T_{b}^{*}(\u^{'})\,\rangle = \delta^{2}_{kr}(\u-\u^{'})\,\cl
\label{eq:c4}
\end{equation}
where $\delta_{kr}$ denotes the Kronecker delta function. This gives us,
\begin{equation}
\cl^{f}=\int_{-\infty}^{\infty}\,\mid \tilde{w}^{\rm eff}(\u-\u^{''}) \mid^{2}\,C_{\ell=2\pi{\rm U}^{''}}\,d^{2}{\rm U}^{''}.
\label{eq:c5}
\end{equation}
Considering that $\cl$ varies much slowly with respect to a faster varying $\mid \tilde{w}^{\rm eff}(\u-\u^{''}) \mid^{2}$ for the $\u$ values being considered \citep{samir14}, $\cl$ can be assumed to be constant over the range, such that,
\begin{equation}
\cl^{f}=\cl\,\,\int_{-\infty}^{\infty}\,\mid \tilde{w}^{\rm eff}(\u-\u^{''}) \mid^{2}\,d^{2}{\rm U}^{''}.
\label{eq:c6}
\end{equation}
The integration yields a value $\int_{-\infty}^{\infty}\,\mid \tilde{w}^{\rm eff}(\u-\u^{''}) \mid^{2}\,d^{2}{\rm U}^{''}=\frac{\pi\theta_{\rm eff}^{2}}{2}$, which results in,
\begin{equation}
\cl=\frac{\cl^{f}}{\frac{\pi\theta_{\rm eff}^{2}}{2}}.
\label{eq:c7}
\end{equation}

We now relate this to the estimated angular power spectrum $\cl^{E}$ which we obtain using the TGE. We note that to estimate $\cl^{E}$ we have assumed,
\begin{equation}
\Delta T_{b}^{f}({\u})=\int_{-\infty}^{\infty}\,{\cal W}(\bm{\theta})\,{\cal A}(\bm{\theta})\,\delta T_{b}^{\rm sky}(\bm{\theta})\,\,e^{+2\pi i \u\cdot\bm{\theta}} d^{2}{\theta}
\label{eq:c8}
\end{equation}
where $\delta T_{b}^{\rm sky}(\bm{\theta})=[{\cal F}(\bm{\theta})\,\delta T_{b}(\bm{\theta})]$. Proceeding as earlier, we see that,
\begin{equation}
\Delta T_{b}^{f}({\u})=\int_{-\infty}^{\infty}\,\tilde{w}^{'{\rm eff}}(\u-\u^{'})\, \Delta T_{b}^{\rm sky}({\u^{'}})\,\,d^{2}{\rm U}^{'}
\label{eq:c9}
\end{equation}
where, $\tilde{w}^{'{\rm eff}}(\u)=F[{\cal W}(\bm{\theta})\,{\cal A}(\bm{\theta})]$ and $\Delta T_{b}^{\rm sky}({\u})=F[\delta T_{b}^{\rm sky}(\bm{\theta})]$, the actual underlying brightness temperature fluctuations from the small radio halo region. Identical to the calculations from eq.~(\ref{eq:c3}) through eq.~(\ref{eq:c6}), we can derive, 
\begin{equation}
\cl^{f}=\cl^{\rm sky}\,\,\int_{-\infty}^{\infty}\,\mid \tilde{w}^{'{\rm eff}}(\u-\u^{''}) \mid^{2}\,d^{2}{\rm U}^{''}
\label{eq:c10}
\end{equation}
and
\begin{equation}
\cl^{\rm sky}=\frac{\cl^{f}}{\frac{\pi\theta_{w}^{'2}}{2}}=\cl^{E}
\label{eq:c11}
\end{equation}
where, $\frac{1}{\theta_{w}^{'2}}=\frac{1}{[f\theta_{0}]^{2}}+\frac{1}{\theta_{0}^{2}}$. Note here in deriving eq.~(\ref{eq:c10}) we have assumed $\langle\,\Delta T_{b}^{\rm sky}(\u)\Delta T_{b}^{*{\rm sky}}(\u^{'})\,\rangle = \delta^{2}_{kr}(\u-\u^{'})\,\cl^{\rm sky}$ with the equivalency that $\cl$ and $\cl^{\rm sky}$ represent the same brightness temperature fluctuations angular power spectrum, i.e. $\cl\equiv\cl^{\rm sky}$. Also note here that our TGE estimator already incorporates a normalization $M_{g}$ which physically captures the effects due to the PB ${\cal A}(\bm{\theta})$ and tapering window ${\cal W}(\bm{\theta})$ in estimated angular power spectrum $\cl^{E}$, which is essentially also captured in the term $\frac{\pi\theta_{w}^{'2}}{2}$ in eq.~(\ref{eq:c11}).

Finally, combining eqs.~(\ref{eq:c7}) and (\ref{eq:c11}), we see that,
\begin{equation}
\cl=\cl^{E}\,\left(\frac{\theta_{w}^{'2}}{\theta_{\rm eff}^{2}}\right).
\label{eq:c12}
\end{equation}
This yields the normalization factor $\left(\frac{\theta_{w}^{'2}}{\theta_{\rm eff}^{2}}\right)$ for sources restricted within a small part of the PB, which is required to estimate the actual $\cl$ from the estimated $\cl^E$ using the TGE. We have next validated this method using simulations in Section \ref{sec:simulation}.

\section{Validating the estimator}
\label{sec:simulation}

\begin{figure*}
\begin{center}
\includegraphics[scale=0.5]{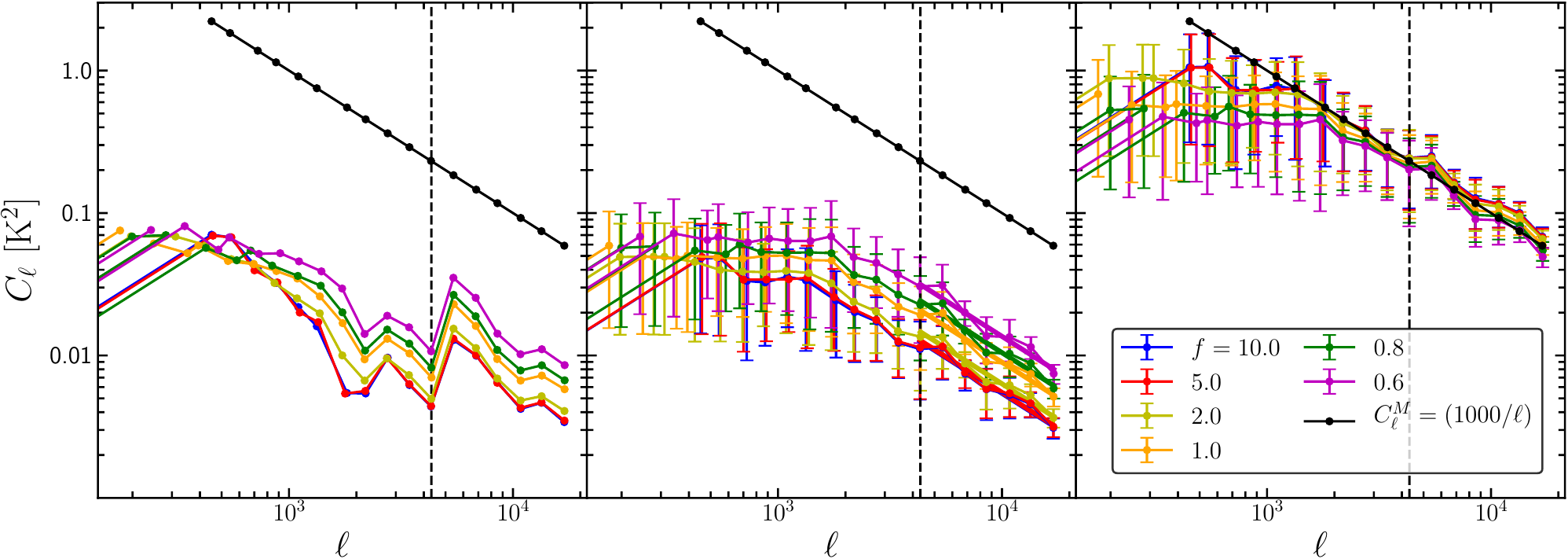}
\caption{$\cl$ as a function of $\ell$ for different tapering values `$f$' as indicated in the legend, considering simulations of Abell 2744 observation. The leftmost panel shows $\cl$ estimated from one realization of the simulations. The data points in the middle panel show $\cl$ with $1\sigma$ error bars estimated from simulations of 10 different sky realizations drawn from the input model angular power spectrum $C^{M}_{\ell}$ (solid black lines in all three panels) at different values of $f$. The solid lines show the best-fit curves for the data points considering a power-law $A(1000/\ell)$, with `$A$' being the parameter being fitted. The rightmost panel shows the same as the middle panel after amplitude correction due to finite size of the galaxy cluster, except the best-fit curves. In all three panels, the vertical black-dashed lines denote the expected $\ell^{\rm GC}_{\rm min}$ values for Abell 2744.}
\label{Sim_1}
\end{center}
\end{figure*}

\begin{figure}
\begin{center}
\includegraphics[scale=0.6]{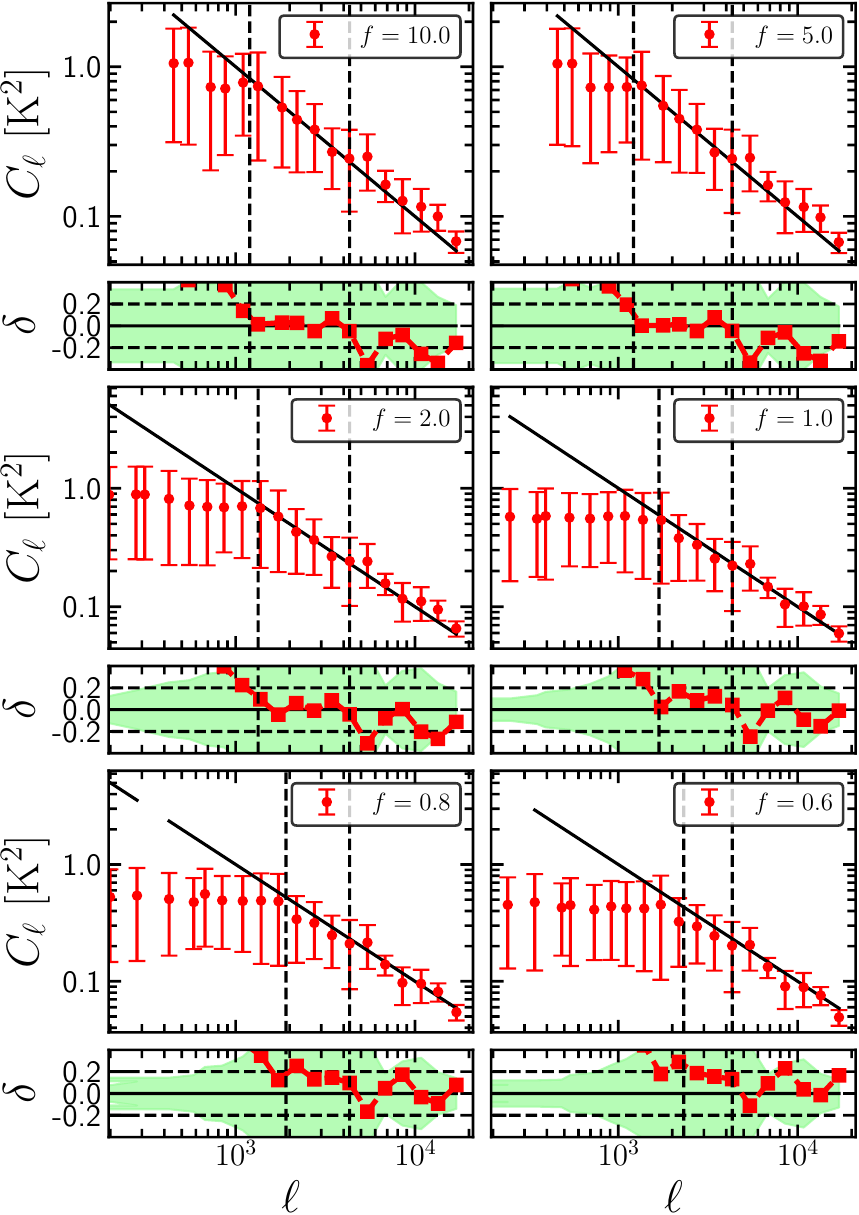}
\caption{Upper parts : The data points show $\cl$ as a function of $\ell$ for simulations of Abell 2744 at different tapering values `$f$' after amplitude correction with $1\sigma$ error bars estimated from simulations of 10 different sky realizations drawn from the input model angular power spectrum $C^{M}_{\ell}$ (black solid lines). Lower parts : The fractional error $\delta = [C^{M}_{\ell} - \cl]/C^{M}_{\ell}$ (data points) and the relative statistical fluctuation $\sigma/C^{M}_{\ell}$ (shaded regions) as a function of $\ell$ for Abell 2744 at different tapering values `$f$' after amplitude correction. In each panel, the left vertical black-dashed lines show $\ell_{\rm min}$ where the convolution in TGE is expected to be important at the particular value of the $f$, whereas, the right vertical black-dashed lines denote $\ell^{\rm GC}_{\rm min}$.}
\label{Sim_2}
\end{center}
\end{figure}

The TGE has already been validated in previous works \citep{samir14,samir16,samir17} using simulations for low-frequency radio observations of all-sky diffuse emissions. In the present work, we have conducted a similar analysis, in which our aim is to study the effect of the finite size of the source on TGE. Note that, so far, 2D TGE has been used to primarily characterize the diffuse emission that fills the entire FoV of the observing telescope; whereas, in the case considered here, we focus on a source restricted within a small part of the PB. We use simulations of the 610 MHz GMRT observation of the galaxy cluster Abell 2744 to establish our methodology for the sources restricted within a small part of the PB.

For the simulations, we have closely followed the prescriptions described in \cite{samir16} and \cite{samir17}. We first generate a wide-field diffuse background sky assuming that the brightness temperature fluctuations $\delta T_{\rm b} (\hat{\bm{n}},\,\nu)$ of the diffuse signal correspond to a Gaussian random field and follow an input  model angular power spectrum given by,
\begin{equation}
    \label{eq:modelaps}
    \cl^{M}  = \left( \frac{1000}{\ell}\right) {\rm K^{2}}\,.
\end{equation}
We have used simulations of $N^{2}=2048 \times 2048$ square grids with grid spacing of $\Delta L \sim 29.54$ Mpc. This corresponds to an angular resolution  of $\Delta\theta \sim 4.98^{''}$ ($\Delta L = r \Delta\theta$, \bm{$r=1223.7\,\,{\rm Mpc}$} being the co-moving distance to the cluster corresponding to the redshift $z=0.308$). The FoV of the simulations ($N\Delta\theta$) covers $\sim 4$ times the $\theta_{\rm FWHM}$ of GMRT at the reference frequency $\nu_c \sim 610 \, {\rm MHz}$. The simulations also cover a bandwidth of $32.94\,\,{\rm MHz}$, centered at $\nu_{c}$, over $22$ spectral channels, where we have considered the source to have a flat spectrum. Next, to obtain the emission corresponding to the galaxy cluster Abell 2744 which is restricted within a small part of the PB, we have multiplied the simulated brightness temperature distribution $\delta T_{\rm b} (\hat{\bm{n}},\,\nu)$ with a Gaussian filter ${\cal F}(\thetavec)=e^{-\theta^{2}/(2\,r_{h})^{2}}$ (defined in Section~\ref{sec3b}), where $r_h$ is the half-light radius of the galaxy cluster in radian. For Abell 2744, $r_h$ is estimated to be around $150^{'}$ \citep{Paul19}. Finally, we have multiplied our sky signal with the GMRT PB ${\cal A}(\th,\,\nu)$ at $\nu_{c}$ and estimated the corresponding visibilities for a baseline distribution identical to the actual Abell 2744 observation considered here. We have used these visibilities to estimate the $\cl$ from the simulated source using the TGE (eq.~\ref{eq:a2}), where we have analyzed the simulated data identical to the actual data.

The leftmost panel of Figure \ref{Sim_1} shows the $\cl$ estimated using the TGE as a function of $\ell$ for different values of the tapering parameter `$f$' for one realization of the simulation. The black solid lines correspond to $C^{M}_{\ell}$ (eq.~\ref{eq:modelaps}), which acts as the reference against which we compare recovered values of the $\cl$ estimated using TGE to validate our estimator. The first thing we notice is that for our source, the amplitude of recovered $\cl$ is suppressed relative to $\cl^{M}$ almost by an order of magnitude at all values of $f$. This contrasts with our previous studies that had considered all-sky diffuse simulations \citep{samir14,samir16,samir17}, where we could recover $\cl^{M}$ with considerable accuracy using the TGE at any value of $f$.  The amplitude suppression in the estimated $\cl$ is observed due to the finite size of the galaxy cluster considered in our simulations. It can be explained by considering the Gaussian filter ${\cal F}(\thetavec)$ introduced over the diffuse background to constrain the cluster size, which causes suppression in power in the $\cl$ domain. We have checked that this is directly related to the extent of the cluster, where if we increase the value of $r_{h}$ in ${\cal F}(\thetavec)$, the suppression of $\cl$ decreases relative to $\cl^{M}$ at a constant value of the tapering parameter `$f$'. Further, the vertical black-dashed lines in all the panels of Figure~\ref{Sim_1} denote the expected $\ell^{\rm GC}_{\rm min}=4320$ for Abell 2744, where $\ell^{\rm GC}_{\rm min} \sim \pi/r_{h}$ ($r_{h}$ in radian) corresponds to the angular scale beyond which we expect to observe the emission from the galaxy cluster. The leftmost panel shows that beyond $\ell > \ell^{\rm GC}_{\rm min}$, the estimated $\cl$ indeed show a power-law behaviour at all values of $f$. To verify the consistency of the power-law relative to $\cl^{M}$, we have used $N_r=10$ independent realizations of the simulations to estimate the mean $C_{\ell}$ and the $1\,\sigma$ errors which we have shown in the middle panel of Figure~\ref{Sim_1} for the different values of `$f$'. We have fitted the mean $\cl$ estimated from $10$ realizations with a power-law of the form $A(1000/\ell)$ at $\ell > \ell^{\rm GC}_{\rm \min}$, where the amplitude `$A$' is the free parameter being fitted. The solid lines in the middle panel of Figure~\ref{Sim_1} show the best-fit curves for the data points considering the power-law $A(1000/\ell)$ at $\ell > \ell^{\rm GC}_{\rm \min}$. We see that in the estimated $\cl$, the slope of the power-law in $\cl^{M}$ is recovered with remarkable accuracy across the $\ell$-range, where the reduced-$\chi^{2}$ for the fits lie between $0.11-0.19$, even though the amplitude is suppressed. This further indicates that we are indeed recovering $\cl^{M}$; the introduction of ${\cal F}(\thetavec)$ that determines the extent of the simulated cluster is causing the suppression in power in the estimated $\cl$, as mentioned earlier. Here we note that in the middle panel, the power-law behaviour extends to even smaller $\ell$ values at $\ell < \ell^{\rm GC}_{\rm min}$, which is not observed in the leftmost panel where we consider only one realization of the simulations. We attribute this to the fact that the mean $\cl$ is estimated by combining measurements from $10$ independent realizations of the simulations, due to which, despite of the presence of ${\cal F}(\thetavec)$, the emissions from the larger angular scales get picked up in the mean $\cl$. Indeed, the extent to which we can estimate the mean $\cl$ in the middle panel is not determined by size of the source $r_{h}$ (equivalently, $\ell^{\rm GC}_{\rm min}$), but rather by the size the of the effective beam width of the telescope after tapering, which is larger than source size, and thus enable us to access $\ell < \ell^{\rm GC}_{\rm min}$. This effect is illustrated better in Figure~\ref{Sim_2}, and we shall discuss this in detail shortly. Here, we only mention that, in reality, we only have access to one realization of the Universe; hence, we have only considered $\ell > \ell^{\rm GC}_{\rm min}$ in our fitting (middle panel of Figure~\ref{Sim_1}), as realistically that is what we have access to (leftmost panel of Figure~\ref{Sim_1}).

Next, we correct for the amplitude suppression in estimated $\cl$ through the normalization derived in Section \ref{sec3b}. We see that the correction factor derived in eq.~(\ref{eq:c12}) depends on the size of the source within the PB, as well as the extent of the effective PB pattern. This is consistent with our observation that the amplitude suppression in estimated $\cl$ is different for different values of `$f$' (leftmost and middle panels of Figure~\ref{Sim_1}). The rightmost panel of Figure~\ref{Sim_1} shows the mean $\cl$ and the $1\,\sigma$ errors obtained from the $10$ realizations of the simulations, after the estimated quantities are corrected for the finite size of the galaxy cluster using the normalization derived in eq.~(\ref{eq:c12}). The black solid line and the vertical black-dashed line correspond to $C^{M}_{\ell}$ and $\ell^{\rm GC}_{\rm min}$ respectively. We find that the amplitude correction (eq.~\ref{eq:c12}) works remarkably well for these simulations and $C_{\ell}$ estimated from the simulations closely match $\cl^{M}$ at all taperings. $\cl^{M}$ lies mostly within the error bars showing the $1\,\sigma$ uncertainty for the estimated $\cl$ values, for the entire $\ell$ range where the diffuse emission is expected to dominate ($\ell>\ell^{\rm GC}_{\rm min}$), and at all tapering values.

The upper parts of the different panels in Figure \ref{Sim_2} show the estimated $\cl$ after the amplitude correction (data points) with corresponding $1\,\sigma$ errors, along with the $C^{M}_{\ell}$, which is shown by the black solid lines in all the panels; $\cl$ at different tapering values of `$f$' are presented in the different panels individually. Considering the tapering window ${\cal W}(\thetavec)$ in our estimator (eq.~\ref{chap2:eq:a1}), previous studies \citep{samir14} have shown that the convolution becomes important at the small $\ell$-values. The $\ell$ modes accessible to us due to this convolution depend on the value of $f$ in ${\cal W}(\thetavec)$ and at $\ell \gtrapprox \ell_{\rm min}= 13.3\sqrt{(1+f^{2})}/(f\times\theta_{\rm FWHM})$ the effect of the convolution can be ignored (see \citealt{samir14} for a detailed discussion). The vertical black-dashed lines in Figure \ref{Sim_2} towards the left show the $\ell_{\rm min}$-values. At lower angular multipoles, for all $f$, we see that the convolution in TGE has become important, as evident from the larger deviations from $\cl^{M}$ at $\ell<\ell_{\rm min}$. Further, this shows that the extent to which we can estimate the mean $\cl$ is not determined by $r_{h}$, but by the size the of the effective beam width of the telescope after tapering, due to averaging for estimating mean $\cl$, even though, as mentioned earlier, only $\ell > \ell^{\rm GC}_{\rm min}$ is relevant for us. The vertical black-dashed lines in Figure \ref{Sim_2} towards the right show the expected $\ell^{\rm GC}_{\rm min}$ for Abell 2744. We clearly see that, at all $f$, $\cl$ is in reasonably good  agreement with $\cl^{M}$ across the entire $\ell > \ell^{\rm GC}_{\rm min}$. The lower parts of each panel of Figure~\ref{Sim_2} show the fractional deviation $\delta = [C^{M}_{\ell} - \cl]/C^{M}_{\ell}$ (data points)  after amplitude correction, and the associated $1\,\sigma$ statistical fluctuations relative to $C^{M}_{\ell}$ (green shaded regions) at different tapering values `$f$' (shown in different panels). Considering $\ell > \ell^{\rm GC}_{\rm min}$, we see that $\mid \delta \mid \lsim\,\, 20 \%$ in most of the $\ell$-bins shown here, at all values of $f$. The deviations in the estimated $\cl$ can be attributed to the uncertainties in the normalization factors $M_{g}$. Note that we have used only $10$ realizations of simulations to estimate the $M_{g}$, which is rather small. Previous studies have shown that $\mid\delta\mid$ decreases if the number of realizations is increased to determine $M_{g}$. Another factor that contributes to the deviations is the poorly sampled baseline distribution of the observed GMRT data. We see that $\delta$ lies within the $\pm 1\sigma$ region for the entire $\ell > \ell^{\rm GC}_{\rm min}$ range at all $f$, which is good considering the low baseline density of the data considered here. This shows that eq.~(\ref{eq:c12}) accurately corrects for the normalization in the TGE for sources restricted within a small part of the PB, and TGE successfully recovers the input $\cl^{M}$ with an accuracy better than $\lsim \,\, 20 \%$ across $\ell \le 4320\, (\ell^{\rm GC}_{\rm min})$ considered here.

To summarize, in this section we validated the TGE for sources with a small finite extent within the PB. Further, Figure \ref{Sim_1} shows that the amplitude of the estimated $\cl$ actually increases with increase in the tapering of the PB of the telescope for such sources (leftmost and middle panels), \textit{i.e.} where the extent of the emission is less than the FoV of the telescope. This is somewhat counter-intuitive, and also differs from the all-sky diffuse emissions, in which case the estimated $\cl$ does not depend on the value of $f$ being considered, and follow $\cl^{M}$ with reasonable accuracy at all $f$. This is again due to the fact that for the sources restricted within a small part of the PB, the normalization becomes $f$-dependent. This is an important observation from these simulations as, as we shall see shortly, this finding will help us in understanding the results obtained from the actual observations (Section \ref{sec3}).

\section{Results}
\label{sec:sec3}

\subsection{Estimated Angular Power Spectrum}
\label{sec:cl}

\begin{figure*}
\begin{center}
\includegraphics[scale=0.6]{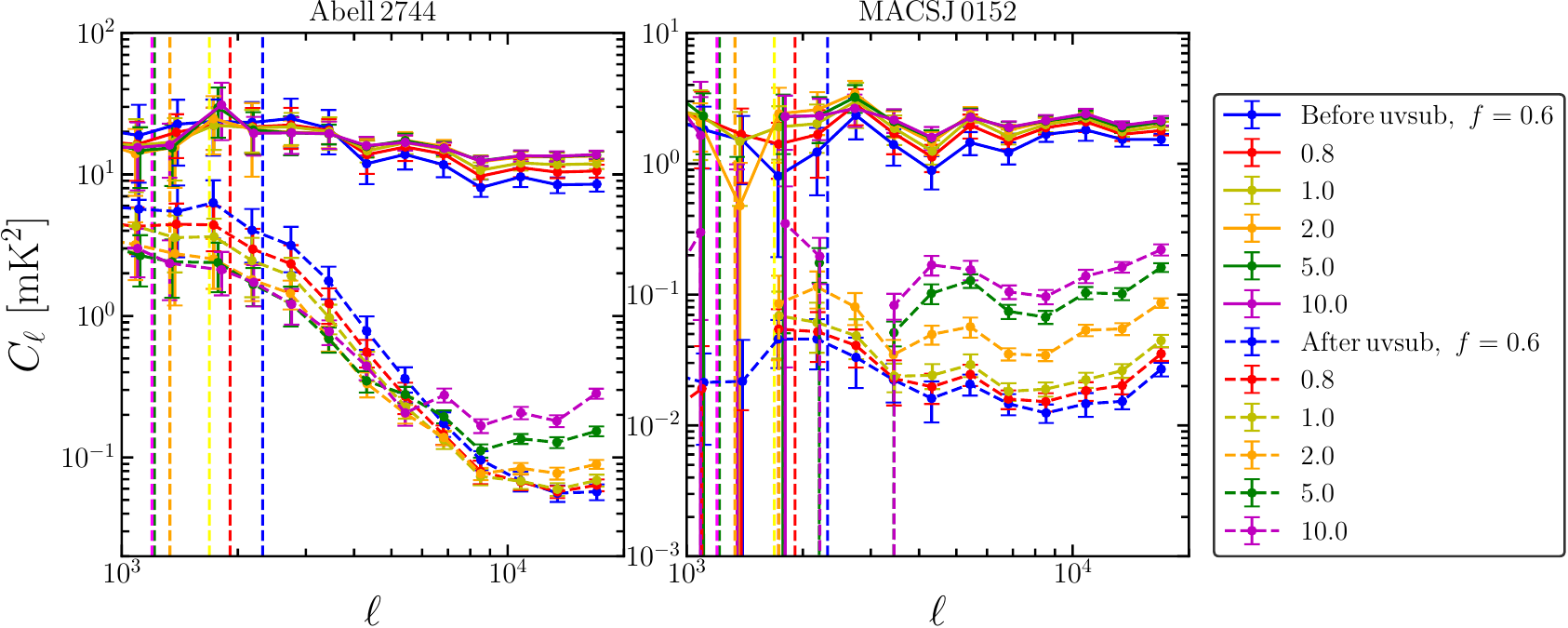}
\caption{$\cl$ as a function of $\ell$ estimated for Abell 2744 (left panel) and MACSJ0152 (right panel) at different tapering values `$f$', as indicated in the legend, before (solid lines) and after (dashed lines) compact source subtraction. The vertical dashed lines show $\ell_{\rm min}$ where the convolution in TGE is expected to be important at the particular value of the $f$ corresponding to the same colour. The $1\sigma$ errors in $\cl$ are estimated through simulations as described in Section 3.3 of \citealt{Saha19}.}
\label{fig:fig4}
\end{center}
\end{figure*}


We have estimated the APS $\cl$ directly from the visibilities using eq. (\ref{eq:a2})  using  the calibrated visibilities both  before and after  point source subtraction. As mentioned earlier, we have used all $22$ frequency channels for a baseline range of $\lvert {\bf U}_{i} \rvert \leq 3000 \lambda$ for the analysis. The Gaussian window function (eq.~\ref{eq:win}) is adopted to taper the GMRT PB pattern and we have considered six values of the tapering parameter `$f$' for this analysis, namely,  $f=0.6,\,0.8,\,1.0\,2.0,\,5.0\, \textrm{and}\,10.0$. Tapering increases with decreasing value of `$f$', and $f=10.0$ is equivalent to an untapered PB pattern. We have generated $10$ realizations of simulations to estimate the normalization factor $M_{g}$. We have binned $\cl$ into $25$ logarithmic bins along $\ell$.

Figure \ref{fig:fig4} shows the binned $\cl$ along with $1\sigma$ error bars as a function of $\ell$ for the galaxy clusters for different values of `$f$' before (shown by the solid lines) and after (shown by the dashed lines) the compact source subtraction. The $1\sigma$ errors consist of the cosmic variance as well as the system noise, and have been estimated through simulations. We have simulated $50$ independent realizations of Gaussian random fields for each of the cases shown in Figure \ref{fig:fig4} such that the fields follow the estimated binned $\cl$ values corresponding to a particular $f$, for a particular source (Abell2744 or MACSJ0152) before or after the point sources have been subtracted. We then estimate the visibilities from these simulated fields using the corresponding GMRT observation with identical baseline, frequency and flagging as the actual data, and add system noise to these estimated visibilities. We find that the the measured visibilities (actual data) have estimated variance of $\sigma^{2}_{N} = (0.28)^{2}\, \textrm{Jy}^{2}\,\,{\rm and}\,\, (0.12)^{2}\, \textrm{Jy}^{2}$ for Abell2744, before and after point source subtraction respectively, and $\sigma^{2}_{N} = (0.15)^{2}\, \textrm{Jy}^{2}\,\,{\rm and}\,\, (0.12)^{2}\, \textrm{Jy}^{2}$ for MACSJ0152, before and after point source subtraction respectively. The system noise contribution to these simulated visibilities are then Gaussian random fluctuations with mean zero and variance $\sigma^{2}_{N}$. We have applied the TGE estimator to the simulated visibilities identical to the actual data for all the cases we have considered. The standard deviation of the estimated $\cl$ obtained from $50$ realizations for each of the case considered gives the estimate of the error due to the cosmic variance plus noise. The vertical dashed lines in Figure \ref{fig:fig4} show $\ell_{\rm min}\sim[2300,\,1900,\,1700,\,1300,\,1200\,{\rm and}\,1100]$ where the convolution in TGE is expected to be important (see Section \ref{sec:simulation}  and \cite{samir14,Pal20} for details) corresponding to $f=[0.6,\,0.8,\,1.0,\,2.0,\,5.0\,{\rm and}\,10.0]$ respectively. Angular scales smaller than this limit (larger $\ell$) give us reliable estimates of the $\cl$.

Before point source subtraction, the estimated $\cl$ are mostly constant across the entire $\ell$ range. At $f=10$, the amplitude is found to be around $20\,\,{\rm mK}^{2}$ for Abell 2744, and $2-3\,\,{\rm mK}^{2}$ for MACSJ0152. The amplitude falls off slightly compared to $f=10$, by a factor of $1.2-1.6$ for Abell 2744 and $1.2-1.9$ for MACSJ0152 at higher $\ell$ considering $f=0.6$, as the tapering in the PB pattern is increased. At $f=0.6$ estimated $\cl$ is $\sim 10\,\,{\rm mK}^{2}$ and $1-2\,\,{\rm mK}^{2}$ in the fields containing Abell 2744 and MACSJ0152, respectively. The foreground is expected to dominate over the radio halo signal at this point. The most significant contributions to the foreground itself come from, (1.) the Poisson part of the compact sources, (2.) the clustered part of the compact sources, and (3.) the diffuse synchrotron background from our own Galaxy, namely DGSE (diffuse Galactic synchrotron emission). At these frequencies, compact sources have the largest contributions, except at the smallest angular multipoles where DGSE may be stronger. Considering $\cl$, the Poisson part of the compact sources is expected to show no variation with $\ell$, whereas, the other two are expected to display distinct $\ell$-dependence that helps distinguish them \citep{ali14,ghosh3,Cha1}. Considering this, Figure \ref{fig:fig4} suggests that estimated $\cl$ before compact source subtraction are dominated by the Poisson part of the compact source contribution (mostly flat along $\ell$), over DGSE and clustering part of the compact source contributions within the fields. Figure \ref{fig:fig1} (middle panels) also shows this; the diffuse emission from the radio halo of Abell 2744 lies below some of the compact sources in the FoV, whereas the halo is not visible for MACSJ0152. The $\cl$ decreases with increase in the tapering in the TGE, but only slightly because the impact of the compact sources within the FoV dominates over the sources in the periphery of the main lobe where the tapering takes place. Note that we have checked that the artifacts seen in the residual images (Figure \ref{fig:fig1}) do not bear much effect on our results.

As the object of our interest is the halo emission, we focus here onward on the results from residual data after subtracting the bright compact sources. We see a large drop in the estimated $\cl$ for both the sources, by a factor of $10-10^2$ or larger (depending on $f$ and $\ell$ values being considered) in the residual data. This further substantiates that the $\cl$ was dominated by the point source contributions before the compact sources were subtracted out of the data. We first focus on Abell 2744 (lower curves of the left panel of Figure \ref{fig:fig4}), where we still observe the contributions from the point sources at higher $\ell$ ($>5000$ at $f=10$), though at a lower level ($\sim 0.2\,\,{\rm mK}^{2}$) compared to before, as indicated by the flat part of the estimated $\cl$ (dashed pink curve in the left panel of Figure \ref{fig:fig4}). We think that the remaining compact source contribution in $\cl$ comes from the unsubtracted point sources within the FoV which are below the flux level of the radio halo being considered for this study (See Figure \ref{fig:fig1} and Section \ref{sec:pssub}). Further, the inaccuracies in modelling the compact sources within the FoV, and the unsubtracted compact sources outside the main lobe of the PB are also expected to contribute to the estimated $\cl$. The contribution from the wide-field compact sources at high angular distances from the phase center of the field are tapered by our estimator, and the $\cl$ amplitude falls as tapering is increased from $f=10$ to $0.6$. At $\ell=17019$, the point source contribution is suppressed by a factor of $\sim 5$ when maximum tapering is applied compared to when there is no tapering. Further, at $f=0.6$, we see that the point source contribution is almost invisible at all the angular scales available to us, with the exception of the highest $\ell$-bin. Considering the lower $\ell$ values, on the other hand, we observe power-law behaviours at all taperings after the point sources have been subtracted. Neglecting the lowest angular multipoles which are dominated by the convolution effect discussed earlier, i.e. $\ell \lsim \ell_{\rm min}$ at different `$f$', we see that the estimated $\cl$ falls gradually through a slope within $1100 \le \ell \le 5427$ at $f=10$, $1726 \le \ell \le 13458$ at $f=0.6$, and values in-between at the intermediate values of `$f$'. The tapering feature in TGE suppress the foreground contribution at highest $\ell$ which allows us to probe the diffuse emissions at larger modes at $f=0.6$. We can further break these $\ell$ ranges into two parts, where each segment follows a distinct and separate power-law (at all values of `$f$'). We discuss this in detail in Section \ref{sec:sec4}. Here, we highlight that within the $\ell$ ranges mentioned above, the $\cl$ amplitude actually increases with the increase in tapering. This behaviour is somewhat counter-intuitive, and contrary to what we observe for the wide-field compact sources. However, our observation is consistent with the results obtained from the simulations (Figure \ref{Sim_1}) discussed in Section \ref{sec:simulation}, which showed that for diffuse emission restricted within a small part of the PB, such as from the halo of the galaxy cluster, the $\cl$ is expected to increase with increase in tapering. This is one of the key findings of this study. This result also indicates to us that the lower angular multipoles displaying the power-laws are dominated directly by the diffuse radio emission from the halo of the Abell 2744. To further confirm this, we have also considered the estimates of the DGSE expected from this FoV, which is expected to dominate the lower $\ell$ after the point sources are removed from the data. We discuss this in Section \ref{DGSE}. In the case of MACSJ0152, the $\cl$ displays power-law at lower $\ell$ (lower curves of the right panel of Figure \ref{fig:fig4}), and contributions from residual compact sources at high $\ell$ (indicated by the flat part of the estimated $\cl$). All point sources above a threshold flux of $0.5\,\,{\rm mJy/beam}$ were indiscriminately modelled and subtracted from the FoV in this case, and we expect that the residual point source contributions are mainly due to unsubtracted point sources including outside the main lobe of the PB, as well as some possible contributions due to point source modelling inaccuracies. Nevertheless, like Abell 2744, the compact source contributions decrease with increase in tapering with a suppression factor of $7-10$ for $f=10\,\,{\rm to}\,\,0.6$ at $\ell > 4000$. Considering lower $\ell$, the power-law are more evident at high tapering ($f\le2$), and it is difficult to comment about the same for $f=10 \,\,{\rm and}\,\, 5$ considering the negative $\cl$ values at $\ell\le4000$. The negative power may be a manifestation of a combination of a few factors, such as poor baseline sampling for convolution in TGE at the two taperings, along with low sky emissions at those angular multipoles. We note that these $\cl$ estimates are not of much concern to us, as they do not hold much bearings for the ensuing study. Considering $\ell\le 4313$, we see overall amplitude suppression at all $\ell$ till $f=2.0$, which saturates at $f=1.0$ considering $862\le\ell\le3446$. The estimated $\cl$ are consistent with each other within $1\sigma$ error bars for $862\le\ell\le3446$ at $f\le1$. This is unlike what is observed in Abell 2744, and is consistent with previous observations and model predictions for all-sky diffuse radiation. In line with the previous studies, $862\le\ell\le 4313$ (at $f=0.6$) are most likely dominated by all-sky diffuse emissions, such as the DGSE, and we check this next.

\subsection{DGSE contribution in the Field}
\label{DGSE}

\begin{figure}
\begin{center}
\includegraphics[width=1.0\columnwidth]{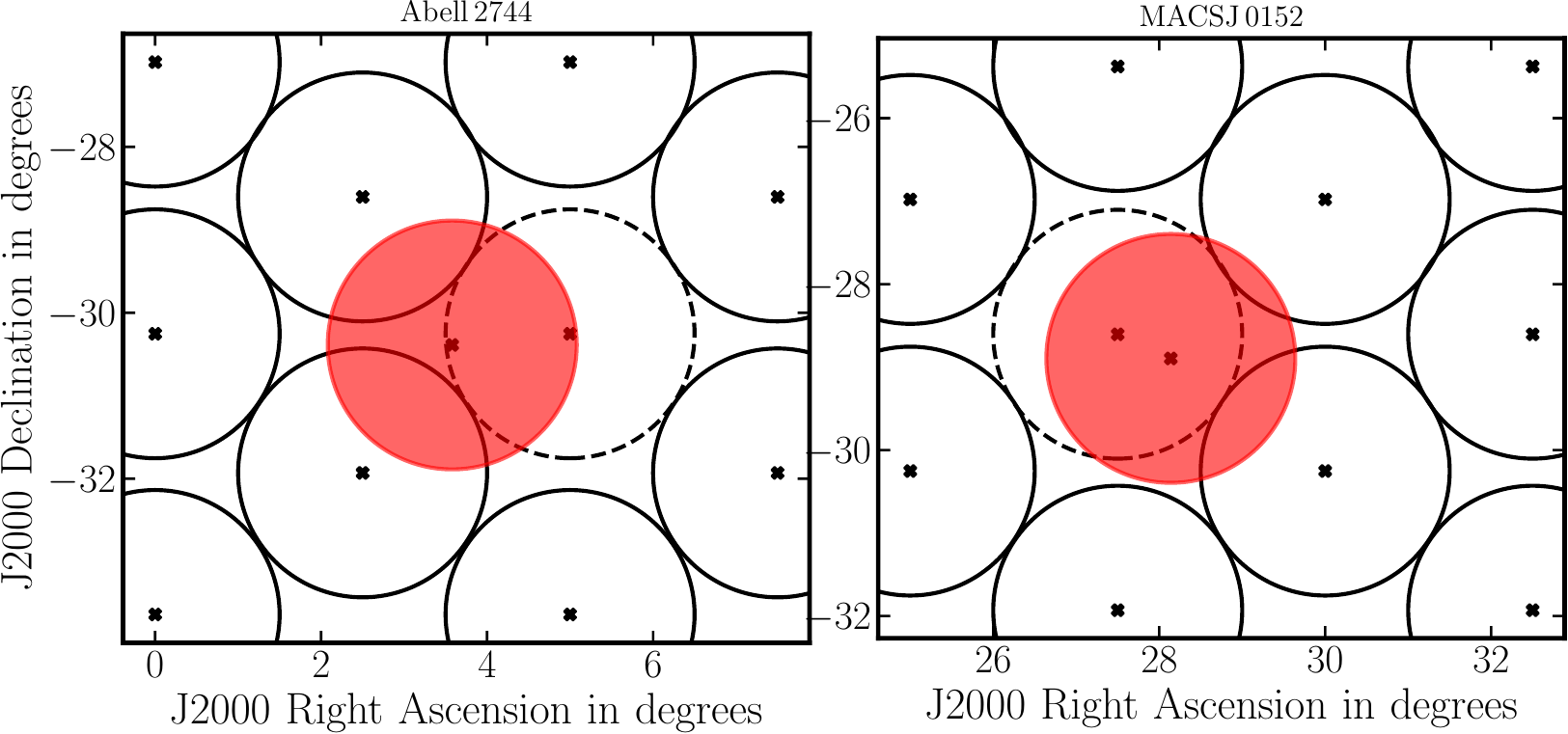}
\caption{This shows the fields surrounding Abell 2744 and MACSJ0152 within a radius of $3.5^{\circ}$ that are observed in the TGSS survey. The black $\boldsymbol{\times}$' denotes the phase centre of the fields, and the black lines around them roughly show the FoV of GMRT for the TGSS survey. The red circular regions show the FoV for our observations at TGSS frequency; the phase centers (`$\boldsymbol{\times}$') correspond to (RA, DEC) of the two galaxy clusters. The dashed circles show the TGSS fields that overlap with our galaxy clusters.}
\label{fig:Surroundingfields}
\end{center}
\end{figure}

\begin{figure*}
\begin{center}
\includegraphics[scale=0.6]{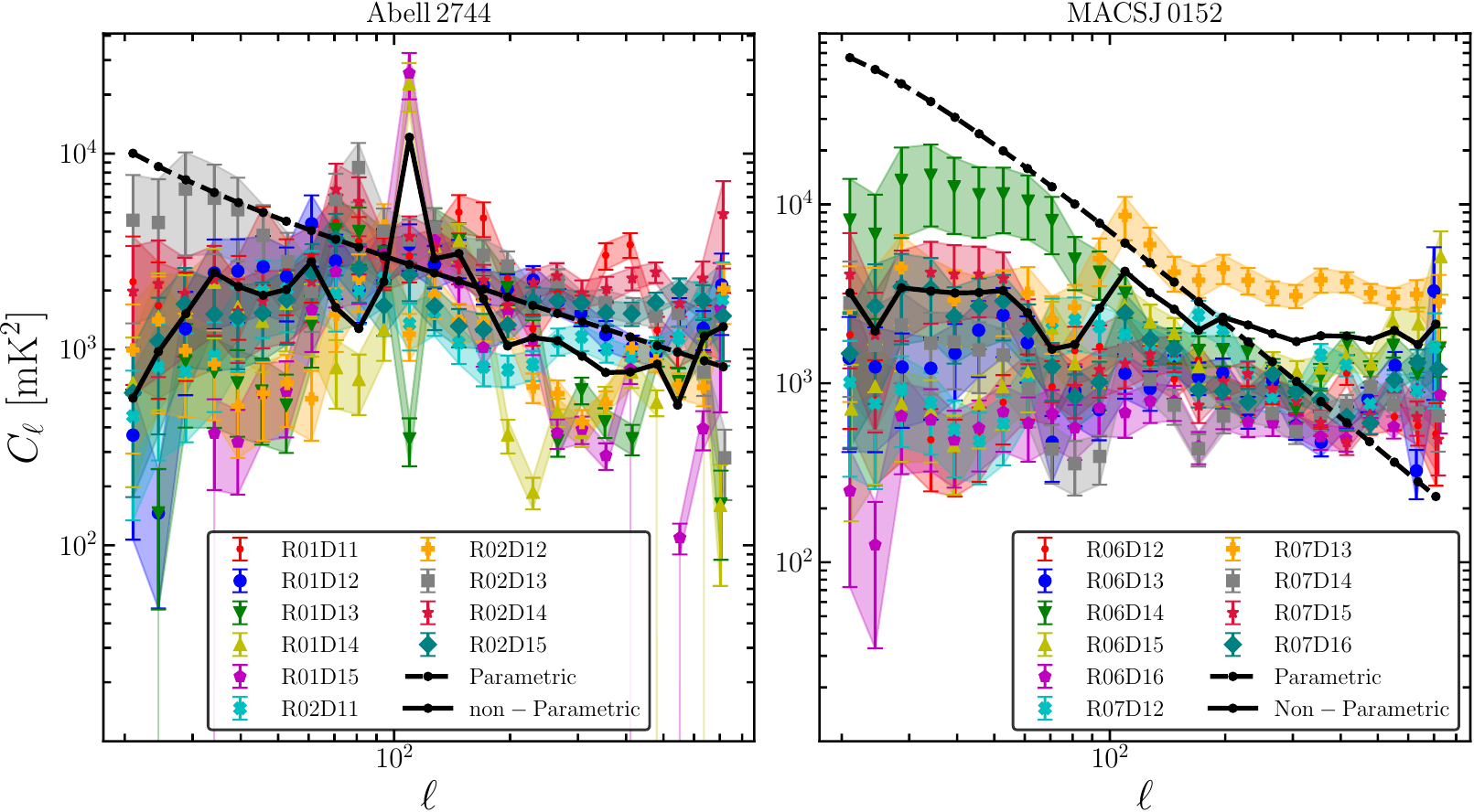}
\caption{The data points show $\cl$ along with $1\sigma$ errors (the shaded regions) estimated by \citealt{samir20} at $\sim147\,\,{\rm MHz}$ for the TGSS fields surrounding Abell 2744 and MACSJ0152, as shown in Figure \ref{fig:Surroundingfields}. The black points show estimates of DGSE expected in our observations as predicted by the Parametric (dashed lines) and non-Parametric methods (solid lines).}
\label{fig:results_fig3}
\end{center}
\end{figure*}

\begin{figure*}
\begin{center}
\includegraphics[scale=0.4]{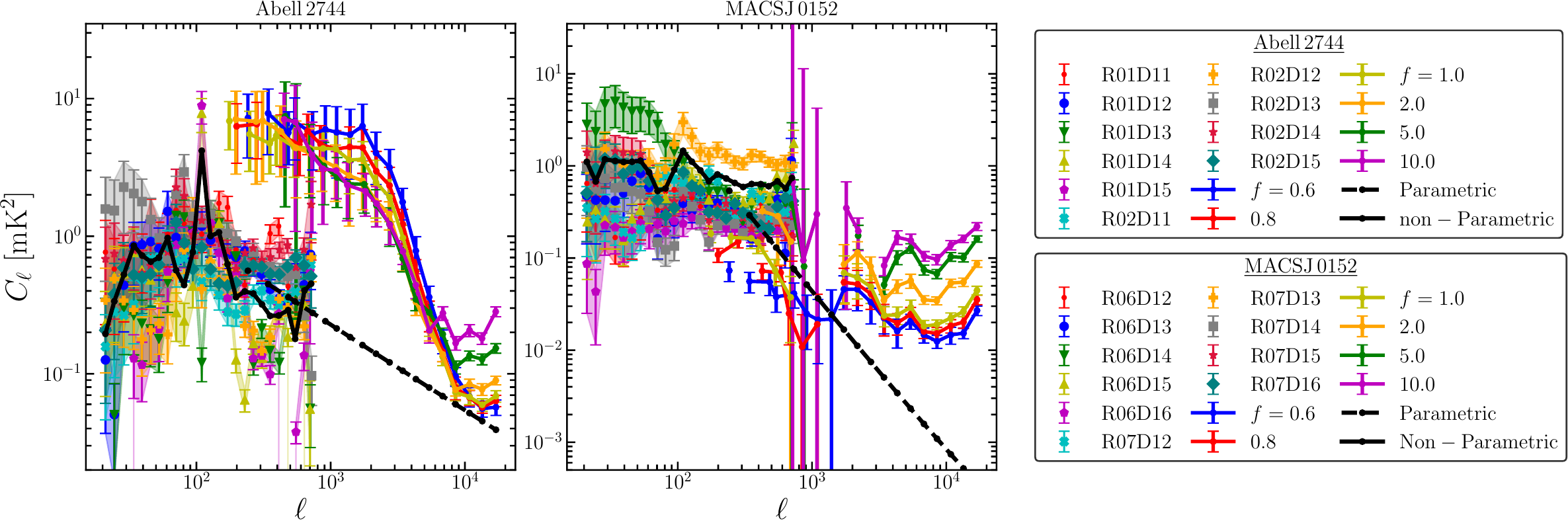}
\caption{The shaded regions along with the data points show the same data as in Figure 6, after the amplitudes are scaled to our observation frequency $\nu_{c}=612.33\, {\rm MHz}$ using a spectral index, $\alpha=2.8$. The Parametric and non-Parametric estimations are also scaled in frequency similarly. The black-dotted solid lines show the non-Parametric predictions at $612.33$ MHz for the DGSE. The black-dotted dashed lines show the expected DGSE contributions in the $\ell$ range for our observations using the Parametric method. The solid lines along with the data points show estimated $\cl$ with $1\sigma$ errors for the clusters at various levels of tapering, after the point source subtraction.}
\label{fig:results_fig4}
\end{center}
\end{figure*}

We next investigate the nature of the diffuse emissions in the two fields. DGSE is a large component of the diffuse emission observed in low frequency radio observations. The DGSE is believed to originate from the interaction between the free electrons in the interstellar medium (ISM) and the Galactic magnetic field \citep{ryb79}. Several observations have characterized the DGSE over a wide range of frequencies, ranging from 10 {\rm MHz} to 100 {\ GHz} \citep{haslam81,haslam82,reich88,deoli08,rema15,zheng17,Irf21}. The statistical properties of the DGSE have been studied in terms of $C_{\ell}$ using all-sky observations \citep{Tegmark96,Tegmark00,Giar01,Giar02,reich01}. The $C_{\ell}$ for DGSE has been shown to be described well by a power-law of the form $C_{\ell}=A(1000/\ell)^{\beta}$ ($A$ is the amplitude) over a large range of scales. The $C_{\ell}$ of the Galactic synchrotron emission measured using single dish all-sky total intensity maps at $408\,\, {\rm MHz}$ \citep{haslam82} and $1420\,\, {\rm MHz}$ \citep{reich82, reich86} have been used to determine the $C_{\ell}$ at angular scales greater than $0.5^{\circ}$ \citep{laporta08}. Considering lower frequencies and large angular length scales, overall the value  of $\beta$ is found to vary in the range $1.8$ to $3.0$ \citep{Bern09,ghosh3,Ia13,samir17,samir20}. The strength (amplitude) of the DGSE ($\sim$ 100s of ${\rm mK}^{2}$), and possibly also the slope $\beta$, are expected to vary across different directions on the sky. \citet{Bern09}, \citet{ghosh3}, and \citet{Ia13} respectively reported $\beta=2.2$ at $l \leq 900$, $\beta=2.34$ at $253 \leq l \leq 800$, and $\beta=1.8$ at $100 \leq l \leq 1300$ using $150\,\,{\rm MHz}$ data of WSRT (Westerbork Synthesis Radio Telescope), GMRT, and LOFAR observations respectively. In a recent work, \citet{Gehlot21} have measured $C_{\ell}$  using $122\,\,{\rm MHz}$ LOFAR-AARTFAAC observations of the NCP, which they find to be consistent with DGSE, with $\beta= 2$ in the range $20\le\ell\le200$. It is essential to characterize the DGSE expected from our fields of interest, in order to distinguish the halo emission from the two galaxy clusters. We have adopted the following method for this endeavour. 

In this work, we have used $\cl$ estimates from \cite{samir20}, who had measured the all-sky angular power spectrum using the TIFR GMRT Sky Survey (TGSS) at 150 MHz, to predict the DGSE contribution for our fields, and at our frequency of observation. For more details on the survey and the methods used to characterize the DGSE, the reader is referred to \citealt{samir20} itself. For our purpose, we first identify the fields surrounding our fields of view that are observed in the TGSS survey. We performed a field search using the TGSSADR Image archive \citep{vo:TGSSADR_images} around the (RA, Dec) of Abell 2744 $(3.58^{\circ},-30.39^{\circ})$ and MACSJ0152 $(28.14^{\circ},-28.90^{\circ})$ with field size $3.5^{\circ}$. This lists $10$ TGSS fields within the radius of $3.5^{\circ}$ around the (RA, Dec) of each of the two galaxy clusters, such that the galaxy cluster overlaps with the $5^{\circ}\times5^{\circ}$ images of the corresponding TGSS fields. Figure \ref{fig:Surroundingfields} depicts a pictorial representation of these fields, where the black `$\boldsymbol{\times}$' denote the phase centre of the fields, with the black lines encircling them representing the $\sim 3^{\circ}$ FWHM of the GMRT PB at $147\,\,{\rm MHz}$ corresponding to the TGSS survey. The red circular regions show the FoV of our observation considering $147\,\,{\rm MHz}$, where the phase centers (presented by `$\boldsymbol{\times}$') are situated at the (RA, DEC) of the two galaxy clusters. Note that the dashed circles in the figure show the TGSS fields that overlap with the galaxy clusters we are considering in the work. These fields are discarded from our analysis, so as not to bias our predicted DGSE estimates with the contribution from the radio halos themselves.

\citealt{samir20} had previously estimated $\cl$ corresponding to 3893 TGSS pointings fields, and had characterized the DGSE using the TGE from the entire TGSS data. We select the fields of our interest, as shown in Figure \ref{fig:Surroundingfields}. The data points along with the shaded regions in Figure \ref{fig:results_fig3} show the estimated $\cl$ along with $1\sigma$ error bars estimated by \citealt{samir20} for our fields at $\sim147\,\,{\rm MHz}$; note that a tapering of $f=1.0$ was used for these results. Here, we have implemented two methods to predict the DGSE for our observations, namely, ``Parametric" and ``non-Parametric". In the Parametric approach, we have identified the $\ell$-range where it appears that the DGSE is dominating, and have performed a fitting on the estimated values of the $\cl$ within this $\ell$-range considering the power-law model of the form $C_{\ell}=A(1000/\ell)^{\beta}$ for each of the individual TGSS fields. The individual DGSE fits for each of the fields are presented in Figure \ref{fig:results_Abell} for both Abell 2744 and Figure \ref{fig:results_MACSJ} for MACSJ0152. Fields R02D13 and R06D14 overlap with our clusters Abell 2744 and MACSJ0152 respectively (shown by dashed circles in Figure \ref{fig:Surroundingfields}), and as mentioned earlier, we have excluded them from the remaining analysis to avoid bias in our DGSE prediction from the galaxy cluster emission itself present in the data. From the remaining data, we have further eliminated two fields for Abell 2744, where the $\beta$ values are found to $-1.910\,\,{\rm and}\,\, 5.547$ after the fitting. We selected these fields based on the criteria that they deviate more than $\delta\beta$ from the mean $\beta$, where $\delta\beta$ is the standard deviation of $\beta$, considering the nine fields surrounding Abell 2744. Following the same criteria, we find four such fields for MACSJ0152 ($\beta=0.638,0.450,2.878,\,\,{\rm and}\,\,3.286$) which we have excluded from further analysis. We have used the DGSE models from the remaining fields ($7$ for Abell 2744, and $5$ for MACSJ0152) to spatially interpolate from the (RA, Dec, DGSE) values of the TGSS fields to the (RA, Dec) of our galaxy clusters. The black-dotted dashed lines in Figure \ref{fig:results_fig3} show the predicted $\cl$ for DGSE in our fields of observations containing Abell 2744, and MACSJ0152 at an frequency of $\sim147\,\,{\rm MHz}$. Figure \ref{fig:results_fig3} shows that our predicted $\cl$ estimates mostly lie within the $\pm1\sigma$ of the estimated $\cl$ for all the fields; the remaining are consistent at $\pm 2\sigma$. The Parametric method is not completely objective in that depending on the $\ell$-range we interpret as DGSE, and the selection criterion for the $\beta$ values, we may get different estimates of the predicted $\cl$ at different $\ell$-values. However, our predicted values lie within $\pm 2\sigma$ of the observed values from the individual fields, which indicates to us that the Parametric method and our DGSE predictions are reliable. The extents of $\cl$ considering all individual fields also give us an idea of the error in our predictions. We have also considered the non-Parametric method, where instead of modelling the DGSE in each fields, we simply interpolate the observed $\cl$ values from the (RA, Dec, estimated $\cl$) values of the nine surrounding fields to the (RA, Dec) of our galaxy clusters. The black-dotted solid lines of Figure \ref{fig:results_fig3} show the predicted $\cl$ for DGSE for Abell 2744 and MACSJ0152 at $\sim147\,\,{\rm MHz}$ using the non-Parametric method. As expected, the non-parametric method predicts $\cl$ completely consistent with all the surrounding fields, irrespective of the value of $\ell$. One drawback in the non-Parametric method, however, is that it can only predict within the observed $\ell$ range, with no way to extrapolate to the observed $\ell$ range measured from the our data.

We next focus on estimating the expected DGSE at our frequency of observation and for the angular scales of our interest. The DGSE has a smooth spectral variation which is usually modelled as $\cl\propto\nu^{-2\alpha}$, $\alpha$ being the spectral index. Observations show that, much like the $\beta$ index, the value of $\alpha$ may vary with depending on the frequency of observation, as well as on which direction on the sky we are observing. In this work, we have considered the mean spectral index of the synchrotron emission $\alpha$ to be $\approx 2.8$ based on \cite{laporta08}. We have used $\alpha=2.8$ to scale the $\cl$ values observed at $\sim147\,\,{\rm MHz}$ for the TGSS survey to obtain the expected $\cl$ values at $\nu_{c}=612.33\,\,{\rm MHz}$. This results in a scaling of the amplitude of all the $\cl$ values estimated from the TGSS survey, as well as the amplitude of the predicted $\cl$ values inferred from them, by a factor of $\approx3.5\times10^{-4}$. The shaded regions along with its data points in Figure \ref{fig:results_fig4} show the same data from Figure \ref{fig:results_fig3} after the amplitude scaling due to the frequency variation. The resulting Parametric and non-Parametric estimations also get scaled by the same amount, and the black-dotted solid lines in Figure \ref{fig:results_fig4} show the non-Parametric predictions at $\nu_{c}=612.33\,\,{\rm MHz}$ for the DGSE, for Abell 2744 and MACSJ0152. As mentioned earlier, the non-Parametric approach is not much useful for predicting the $\cl$ at our angular scales of interest. The solid lines along with the data points in Figure \ref{fig:results_fig4} show the estimated $\cl$ with $1\sigma$ errors for Abell 2744 and MACSJ0152 at various levels of tapering, and after the point source subtraction. We use the parametric method discussed above, along with the frequency scaling, to predict the $\cl$ for DGSE at all the $\ell$ values from our observation. The black-dotted dashed lines show the expected DGSE contributions in our observations and in our $\cl$ estimates for the radio halos of Abell 2744 and MACSJ0152, as obtained using the parametric method. Considering Abell 2744, the estimated $\cl$ are above the DGSE at all $\ell$ for all values of $f$. For MACSJ0152, the estimated $\cl$ lie above the DGSE at all the $\ell$ values where we have meaningful $\cl$ estimates ($\ell \gsim 1700$) at all tapering. Considering even the highest tapering, the residual point source contributions dominate significantly at smaller angular scales ($\ell>4000$), whereas at the larger angular scales $1700\lsim\ell\lsim4000$, the $\cl$ is consistent with the DGSE predictions within the $2\sigma$ errors. This is true even a lower tapering values (barring $f=10\,\,{\rm and}\,\,5$) except that the $\ell$ ranges shift slightly. Considering the non-Parametric approach (solid black lines), for Abell 2744, one may expect the DGSE to be at a similar level as the parametric counterparts; however, nothing more meaningful may be inferred from the non-Parametric approach, especially for MACSJ0152. Figure \ref{fig:results_fig4} shows us that the low $\ell$ for Abell 2744 field are dominated by the emissions from the radio halo of Abell 2744, which is the brightest source in the field after the compact sources have been subtracted, and emission from the halo is much larger than the DGSE in the fields, even almost by an order at the lowest few $\ell$ values. Based on the flatness of the estimated $\cl$ and the tapering pattern in the highest $\ell$ modes, we think that these modes are dominated by the point source contributions, as discussed previously for the total data. For MACSJ0152, the diffuse emissions at the lower $\ell$ are in agreement with the expected DGSE emission (especially, at $f\le2$), whereas the higher $\ell$ and lower tapering ($f>2$) are largely dominated by the point sources.

Next, we discuss some sources of uncertainty in our DGSE estimates; (1.) As mentioned earlier, the Parametric method, though robust, may be subjective at times and we try to account for that our best in the analysis by maintaining uniform criterion when we consider which fields are meaningful in our analysis, as well as by considering the non-Parametric method; (2.) We have used $\alpha=2.8$ following \citealt{laporta08} to account for the spectral variation in $\cl$. However, we note that other studies may suggest slightly different values for $\alpha$ depending of the direction and frequency of the observations. For example, \citet{laporta08} itself had reported $\alpha$ to vary between $2.8-3.2$ for DGSE observations at $408\,\,{\rm MHz}$ and $1420\,\,{\rm MHz}$, which had showed that $\alpha$ may vary significantly depending on how far away from the Galactic plane one is observing. \citet{rogers08} constrained the mean spectral index of the synchrotron emission $\alpha$ to be $\approx 2.5$ at high Galactic latitude and in the frequency range $150 - 408 \,\, \rm MHz$. Such variation in $\alpha$ values may change the DGSE values by a few factors than what we see in Figure \ref{fig:results_fig4}, but we have checked that this will not affect our conclusions about the origin of the emissions; (3.) Further, our DGSE estimations are also restricted by the limitations of the TGSS observations and the analysis performed in \citealt{samir20}, whose results we have used in this work. However, these DGSE estimates are adequate for this study, and the uncertainties discussed here will not affect our conclusions, as discussed already.

\section{Discussions}
\label{sec:sec4}

\begin{figure}
\begin{center}
\includegraphics[scale=0.45]{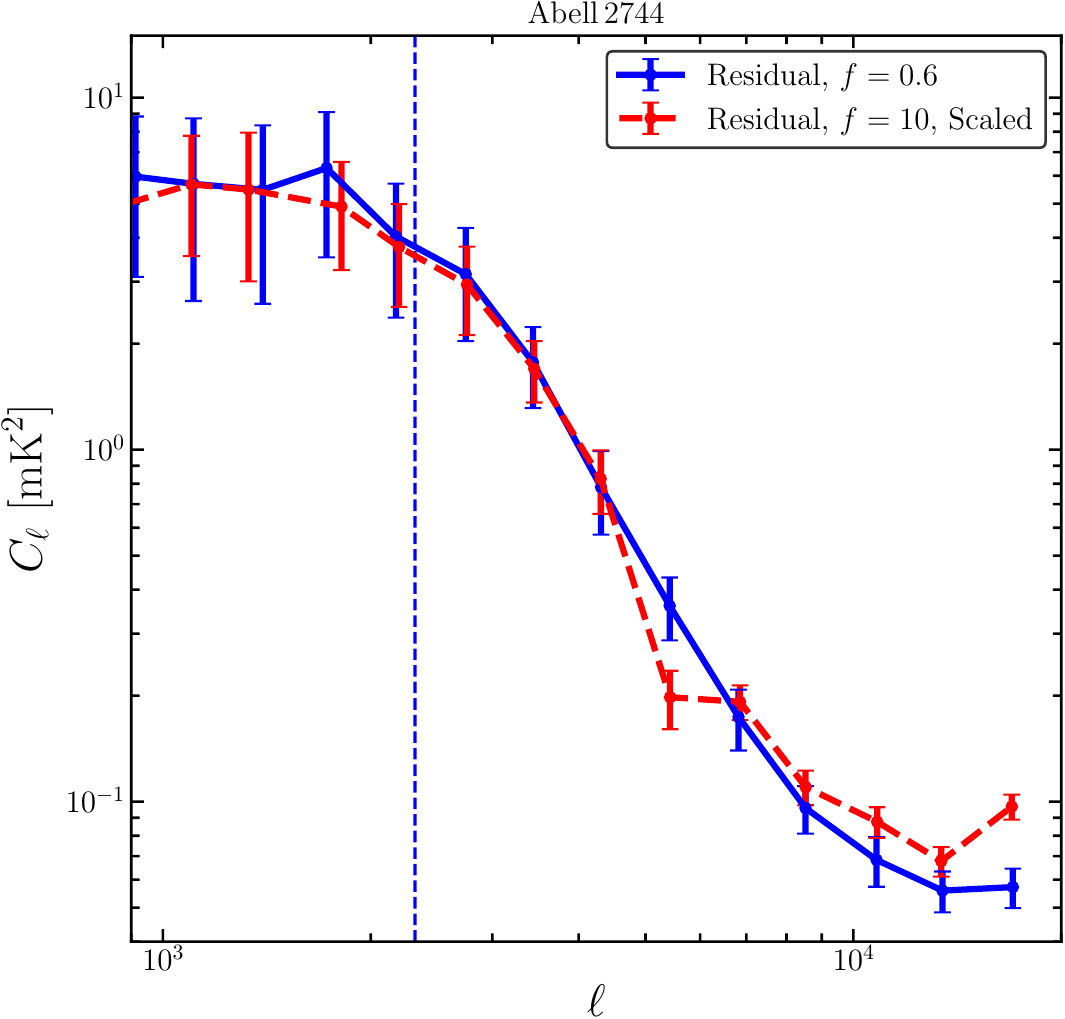}
\caption{The observed $\cl$ at $f=0.6$ (blue-solid lines with data points) for the residual data for Abell 2744 are shown with $1\sigma$ errors. The same is also shown at $f=10$ (red-dashed lines with data points) after we have scaled the estimated $\cl$ to account for the small and finite size of the radio halo. The vertical blue-dashed line denotes $\ell_{\rm min}$.}
\label{fig:fig_scaled}
\end{center}
\end{figure}

\begin{figure}
\begin{center}
\includegraphics[width=\columnwidth]{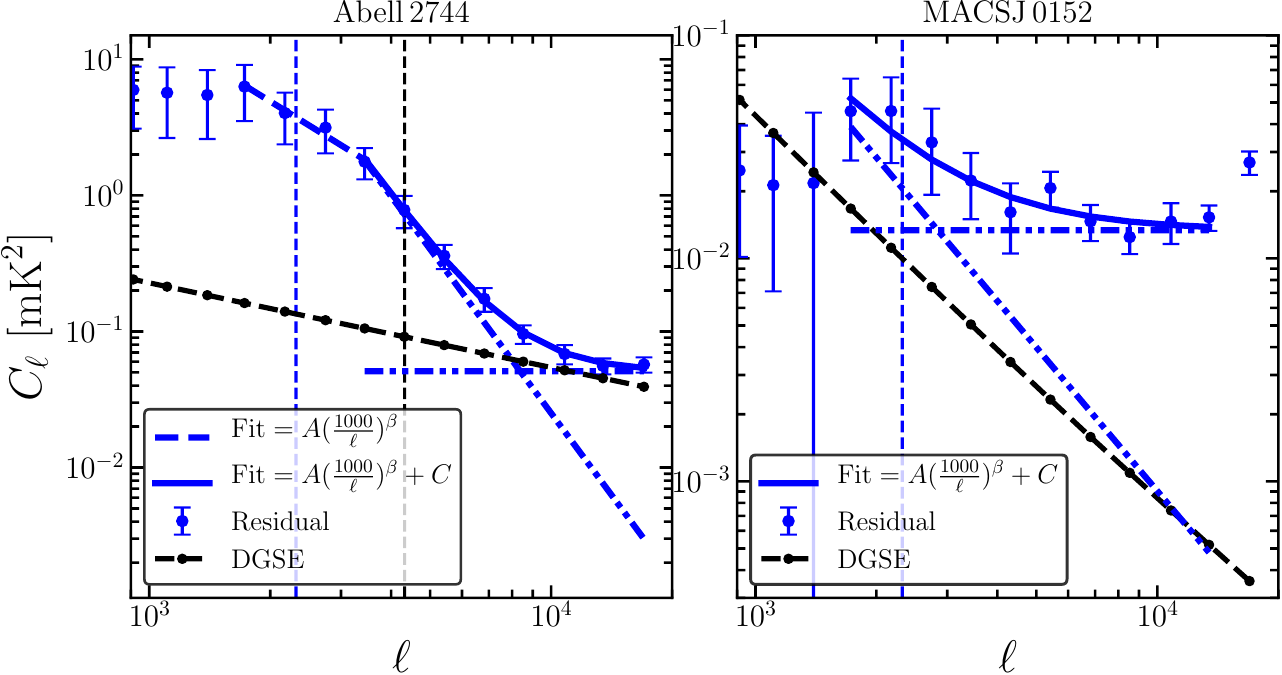}
\caption{Piecewise power law model fit to the $\cl$ for the residual data. The observed $\cl$ (blue data points) at $f=0.6$ for the residual data are shown with $1\sigma$ errors along with estimated DGSE contribution (black-dashed lines). The best-fit models are also shown. The vertical blue-dashed lines denote $\ell_{\rm min}$; the vertical black-dashed line in the left panel denote the $\ell$ corresponding to $300^{''}$ angular size of the radio halo of Abell 2744.}
\label{fig:fig3a}
\end{center}
\end{figure}

\begin{table*}
    \centering
    \caption{The  best fit parameters for a piecewise power law model given by $\cl={\rm A}\big(\frac{1000}{\ell}\big)^{\beta}$ or $\cl={\rm A}\big(\frac{1000}{\ell}\big)^{\beta}+C$.}
    \label{tab:tab_1}
    \begin{tabular}{cccccc}
    \hline
    \hline
      & ($\ell_{\rm min}, \ell_{\rm max}$) & $A\,\,[{\rm mK}^{2}]$ & $\beta$ & $C\,\,[{\rm mK}^{2}]$ & ${\rm Reduced}-\chi^{2}$ \\
    \hline
    Abell 2744 & ($1726, 3436$) & $17 \pm 3$ & $1.8 \pm 0.2$ & $-$  & $0.08$ \\
    & ($3436, 17019$) & $253 \pm 57$ & $4.0 \pm 0.1$ & $ 0.051 \pm 0.002$  & $0.10$ \\
    \hline
    MACSJ0152 & ($1726, 13449$) & $0.12 \pm 0.08$ & $2.1 \pm 0.7$ & $0.013 \pm 0.002$  & $0.52$ \\
    \hline
    \hline
    \end{tabular}
\end{table*}

\begin{figure}
\begin{center}
\includegraphics[scale=0.4]{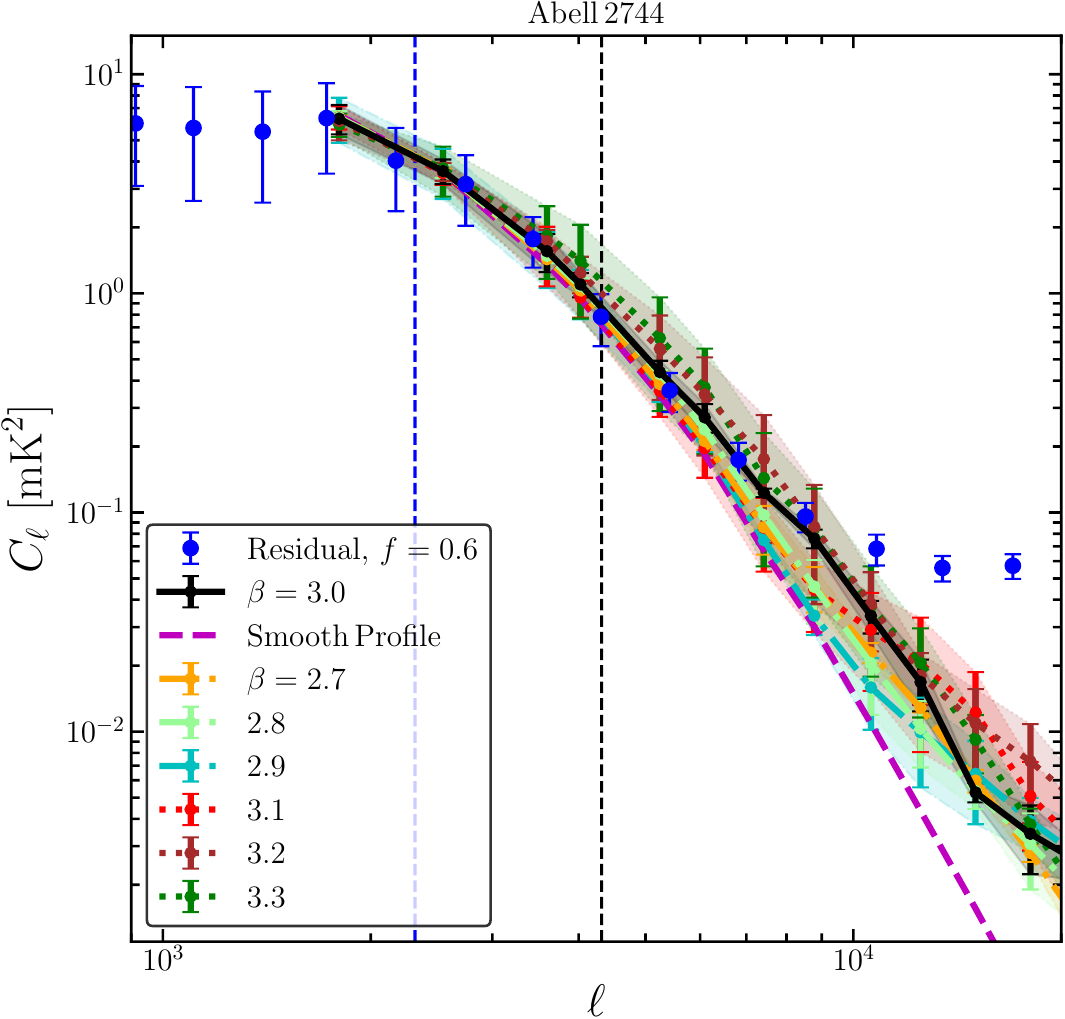}
\caption{The observed $\cl$ at $f=0.6$ (blue data points) for the residual data for Abell 2744 are shown with $1\sigma$ errors. The recovered $\cl$ with $1\sigma$ errors estimated from three independent realizations of the simulated model with an exponential radial profile and power-law fluctuations for the surface brightness are also shown. The model with $\cl\propto\ell^{-3}$ power-law fluctuations (black-solid line and data points) agrees with the observed $\cl$.  We have also shown the scenario where the halo only has a smooth exponential profile with no added fluctuations (magenta-dashed line). Additionally, the expected $\cl$ for six representative values of $\beta$ are shown by the dashed-dotted ($\beta<3$) and dotted lines ($\beta>3$) where the power-law fluctuations are denoted as $\cl\propto\ell^{-\beta}$ in the surface brightness model; the shaded regions show the $1\sigma$ cosmic variance estimated from three independent realizations of the simulations. The vertical blue-dashed line denotes $\ell_{\rm min}$; the vertical black-dashed line denotes the $\ell$ corresponding to $300^{''}$ angular size of the radio halo of Abell 2744.}
\label{fig:fig3}
\end{center}
\end{figure}


For the remaining analysis, we consider only $f=0.6$ from the residual data, which represent the best data available at this stage. First, we consider Abell 2744 for which we could measure $\cl$ from its halo which has a smaller, finite size within the PB. We have checked if the measured $\cl$ for Abell 2744 at $f=0.6\,{\rm and}\,10$ are consistent with each other like we find in Section \ref{sec:simulation}. We note that the factor derived in Section \ref{sec3b} and used in Section \ref{sec:simulation} assume that the underlying field follows a Gaussian distribution. However for Abell 2744 we find that small Gaussian fluctuations having power-law $\cl$ over a smooth brightness profile, both of which having radial exponential variations reproduces the $\cl$ at $f=0.6$; we discuss this in detail shortly. Considering such non-Gaussian surface brightness distribution, eq. (\ref{eq:c12}) needs appropriate modifications to generalize. We have achieved that by redefining $\theta_{\rm eff}$ from eq. (\ref{eq:c7}) for any field , $\frac{1}{\theta_{\rm eff}^{2}}=\frac{1}{m\theta_{1}^{2}}+\frac{1}{[f\theta_{0}]^{2}}+\frac{1}{\theta_{0}^{2}}$ where $m$ is a factor such that $m\theta_{1}^{2}$ approximately represent the square of a characteristic length for the underlying field which we assume can be represented by $e^{-\frac{\theta^{2}}{m\theta_{1}^{2}}}$. Considering measured $\cl$ at any two tapering parameters $f_{1}$ and $f_{2}$, we can estimate the factor $m\theta_{1}^{2}$ using eq. \ref{eq:c12} from observed data, which we can transfer to other $f$ values. Note that here we need no pre-assumptions regarding the halo size since the low-level emissions can extend to much larger length scales as indicated in recent observations. We have used the $\cl$ estimated at $f_{1}\,{\rm and}\,f_{2}=5\,{\rm and}\,0.8$ respectively (Figure \ref{fig:fig4}) to find $m\theta_{1}^{2}$, which we have used to find the relative scaling between $f=10\,{\rm and}\,0.6$ (eq. \ref{eq:c12}). The dashed-red line in Figure \ref{fig:fig_scaled} show the scaled $\cl$ for $f=10$ which we find is very well consistent with the estimated $\cl$ at $f=0.6$ shown in the blue-solid line with data points.

The residual $\cl$ at $f=0.6$ is shown in Figure \ref{fig:fig3a} for both sources. After compact source subtraction, effect of residual compact sources remain at the highest $\ell$, as indicated by the flat part of the estimated $\cl$ (highest two $\ell$-bins for Abell 2744, $\ell>4000$ for MACSJ0152). The lower $\ell$ values, on the other hand, display power-law behaviours; we see a broken power-law for Abell 2744. The radio halo of Abell 2744 has an estimated angular size of $\sim5^{'}$ (${\rm i.e.\,\, physical\,\,length\,\,scale,\,\, L}\sim1390\,{\rm kpc}$) at $610\,\,{\rm MHz}$ \citep{Paul19}, and is expected to contribute at $\ell\gsim4320$ (vertical black-dashed line in Figure \ref{fig:fig3a}). The black dots with the dashed lines in Figure \ref{fig:fig3a} show the predicted DGSE for the centre of the two observed fields, at the locations of the galaxy clusters. The residual $\cl$ is significantly larger (more than an order of magnitude at $\ell=1726$) than the predicted DGSE for Abell 2744 at all angular scales; for MACSJ0152, the DGSE is mostly consistent with the residual $\cl$ at $2\sigma$ level within $1726\le\ell\le 4313$, where the diffuse emission is expected to dominate. The excess over DGSE is likely from the halo emission for Abell 2744, whereas, for MACSJ0152 the diffuse emission is most likely dominated by the DGSE. Finally, we have checked that, after compact source subtraction, the expected point source contribution due to the Poisson fluctuations \citep{ali14} are also consistent with the flattening at large $\ell$ in our observation.

The non-detection of excess non-thermal synchrotron emission for MACSJ0152, as compared to Abell 2744, may be attributed to the dynamical states of the two clusters. Abell 2744 is a markedly disturbed system undergoing a sequence of complex merger events that have profoundly re-shaped its ICM. Gravitational-lensing and X-ray studies reveal at least four interacting subclusters, with pronounced offsets between the dark-matter, galaxy, and gas components, indicative of multiple post–core-passage collisions \citep{Kemp04,Mert11,Owers11}. These interactions supposed to have generated widespread shock heating and turbulence throughout the central region, providing the physical conditions expected to sustain cluster-scale synchrotron emission. Indeed, the diffuse radio halo observed in Abell 2744 is consistent with turbulence driven by its highly dynamical and ongoing assembly history \citep{Pea17,Paul19}. \cite{Vel25} and \cite{Yu22} describe that the cluster MACSJ0152 to have intermediate dynamical state based on different dynamical state proxies that utilize optical and X-ray imaging data. Using the concentration parameter, \cite{Ro17} further identifies the cluster as hosting a cool core, consistent with a relatively undisturbed central ICM. Its other global properties i.e. mass, X-ray luminosity, and temperature are also comparatively modest when set against more massive, violently disturbed systems such as Abell 2744 \citep{Ebe10,Man16}. Taken together, these indicators suggest that MACSJ0152 likely harbours only low-level turbulence, insufficient to generate a detectable radio halo at the sensitivity of our observations, which causes non-detection of excess power in $\cl$.

Following examples from literature \citep[e.g.][and references therein]{Tegmark96,Giar01,ali14,Saha19,Paul19}, phenomenological (broken) power law fits to the estimated $\cl$ for both sources are presented here. The measured $\cl$ are modelled using a simple power-law of the form $\cl={\rm A}\big(\frac{1000}{\ell}\big)^\beta$ (Model 1), `${\rm A}$' being the amplitude of the power spectrum and $\beta$ being the power-law index. The parameters $({\rm A},\beta)$ capture the amplitude of the distribution of the specific intensity fluctuations in the sky and its variation across different angular scales respectively. The measured $\cl$ from the radio halo of Abell 2744 (Figure \ref{fig:fig3a}) clearly shows a break in the power-law at $\ell=3436\,\,(\theta\sim6.3^{'}\,\,{\rm or},\,\,{\rm L}\sim\,1755\,{\rm kpc\,in\,\,physical\,\,length\,\,scale})$. We have modelled $1726\le\ell\le3436$ ($1755\le{\rm L}\le3500$ kpc) with Model 1 and $\ell \ge3436$ (${\rm L}\le1755$ kpc) is modelled using $\cl={\rm A}\big(\frac{1000}{\ell}\big)^\beta+C$ (Model 2), where `$C$' is a constant and gives the contribution due to the unsubtracted point sources \citep{ali14}. Considering MACSJ0152, we could model the entire angular range considered with Model 2. The smallest $\ell$ values are dominated by the convolution due to the primary beam of the telescope and we have not considered them in modelling. The best-fit values for the parameters and the errors are estimated using non-linear least squares; the details of the modelling and estimated best-fit parameters for $({\rm A},\beta,C)$ along with their $1\sigma$ error, and the reduced-$\chi^{2}$ are listed in Table \ref{tab:tab_1}. The reduced-$\chi^{2}$ values for Abell 2744 are quite low, and may indicate over-fitting, or somewhat overestimation of the errors. The errors in $\cl$ in Figure \ref{fig:fig3a} are estimated through simulations following Section 3.3 of \cite{Saha19} under the assumption that the underlying emission originates from Gaussian random field; they may change where the assumption does not hold. We have shown the estimated $\cl$ for the two fields in Figure \ref{fig:fig3a}, along with their best-fit models (dashed- and solid-blue lines for Model 1 and Model 2, respectively), and $\ell_{\min}$ (vertical blue-dashed lines). The radio halo of Abell 2744 has an estimated angular size of $\sim5^{'}$ (${\rm L}\sim1390\,{\rm kpc}$) at $610\,\,{\rm MHz}$ \citep{Paul19}, and is expected to contribute at $\ell\gsim4320$ (vertical black-dashed line in Figure \ref{fig:fig3a}). The non-thermal synchrotron emission from the radio halo has a measured amplitude and slope of $({\rm A},\beta)=(253 \pm 57 \,\,{\rm mK}^{2},\,\,4.0 \pm 0.1)$. The amplitude falls to $17 \pm 3\,\,{\rm mK}^{2}$ at larger angular scales with shallower dependence on $\ell$. $({\rm A},\beta)$ values of the fitted power-law (Table \ref{tab:tab_1}) indicate that the non-detection of the synchrotron emission from MACSJ0152 is consistent with DGSE within $1\sigma$ errors.

Henceforth, we focus on the excess halo emission from Abell 2744. To reproduce the observed $\cl$, we have also simulated the halo emission for Abell 2744 with a radial profile $f(r)=I_{0}\,e^{-ar}$. We find $I_{0}\sim4\,\,{\rm mJy/beam}$ and $a\sim0.016\,\,{\rm arcsec}^{-1}\,({\rm L}\sim298\,\,{\rm kpc})$ from the best-fit exponential model for the observed azimuthally averaged surface brightness. However, a smooth exponential profile, by itself, could not reproduce the observed $\cl$. The magenta-dashed lines in Figure \ref{fig:fig3} show the expected $\cl$ from a smooth exponential; this gives a $\chi^{2}$ value of $22.59$ which corresponds to a
$p-{\rm value} < 0.01$ and reduced$-\chi^{2} \approx 3.28$. Note that these statistics are obtained using $\cl$ within $1700 \lsim\ell \lsim 8525$ as the highest $\ell$-bins are dominated by residual point sources. Hence, on top of the exponential profile, we introduce small fluctuations $(1+\delta)$ in our simulation as a multiplicative factor in $f(r)$. If we take the fluctuations $\delta$ to be zero mean Gaussian random field that has an underlying power spectrum $\cl\propto\big(\frac{1000}{\ell}\big)^{3.0\pm0.1}$, the APS from the simulation is found to be consistent with the observed $\cl$. The black-solid line and data points in Figure \ref{fig:fig3} shows the recovered mean $\cl$ with $1\sigma$ standard deviations estimated from three independent realizations of the simulations for $\cl\propto\ell^{-\beta}$ for $\beta=3.0$ for the fluctuations in our model. Note that the amplitude of $\cl$ here is not of our interest and is treated as a free parameter in the simulations. We have considered the reduced$-\chi^{2}$ values to determine best-fit $\beta$ from the theoretical and observed $C_{\ell}$ estimates within $1700 < \ell < 8525$. We have used simulations for $2.4 \le \beta \le 3.3$, for which we estimated the reduced$-\chi^{2}$. We find that for $\beta = 3.0$, the reduced$-\chi^{2} \approx 0.22$ is the minimum and we could recover the estimated $\cl$ (blue data points) accurately at the lowest $\ell$-values. The theoretical $\cl$ with $1\sigma$ errors (shaded regions) for six additional representative values of $\beta$ are also shown for comparison. The reduced$-\chi^{2}<1$ for $\beta = 3.0$ may be due to over-fitting or over-estimation of the errors. The reduced$-\chi^{2}$ increases significantly at both $\beta=2.9\,{\rm and}\,3.1$ and at all $\beta$ values considered (excluding $\beta=3.0$) are much greater than one which shows poor fit. Based on our analysis we report $(1000/\ell)^{3.0 \pm 0.1}$ as the best-fit model which best recovers the observed $C_{\ell}$. Deviations at the higher $\ell$ are most likely due to non-inclusion of residual compact sources in the simulation that are present in the data. The power-law APS of the specific intensity fluctuations of the non-thermal synchrotron emission from the observed radio halo of Abell 2744 likely arises from turbulence in its ICM.

The observed power-law power spectrum ($\cl\propto\ell^{-3}$) can be compared with different theoretical turbulence models to investigate the nature of the turbulence in Abell 2744. Similar analysis have been performed by \citealt{chu12} which used the 2-dimensional power spectrum to study 3-dimensional turbulence in the ICM using X-ray data. There are however additional complications to adapt similar methods to compare models of 3-dimensional turbulence for the radio data. The X-ray emission from the ICM is related to turbulence only through the underlying density. In radio frequencies, the synchrotron emissions coming from ICM turbulence, however, additionally depends on the underlying magnetic field which introduces an additional dependency. The non-thermal synchrotron specific intensity fluctuations are expected to follow the underlying magnetic field and electron density fluctuations. The synchrotron emissivity relates to electron density, $n_{e}$, and two-dimensional magnetic field perpendicular to the line-of-sight, $B_{\perp}$ as, $\propto n_{e}\times\vert B_{\perp}\vert^{(p+1)/2}$, where $\vert x \vert$ denotes the magnitude of the vector field `$\bm{x}$', and `$p$' is the power-law index for the energy distribution of the cosmic-ray electrons, and is related to the radio spectral index ($\alpha$) as, $p=1-2\alpha$. Depending on the correlation between distribution of $n_{e}$ and the underlying fluctuations in the magnetic field in the ICM, relating the $\cl$ and the magnetic field power spectrum becomes complicated. The actual nature of correlation between $n_{e}$ and the magnetic field in the ICM is unclear; different scenarios include (1.) $n_{e}$ shows a strong correlation with the underlying fluctuations in the magnetic field, (2.) the cosmic-ray electron population is nearly constant and independent of the magnetic field or (3.) the intensity fluctuations are then related to the magnetic field fluctuations only. A detailed comparison of the measured $\cl$ with theories requires full 3-dimensional simulations of ICM turbulence from first principle, which we note is beyond the scope of the present work. A thorough comparison between observations and these models will be addressed in a future work (T. Sharma et al. 2026, in preparation).

Our work also shows that $\cl$ estimation is a useful tool to study radio halo structure beyond the smooth exponential surface brightness profile. This second order statistics integrates the weak emissions over the large volume of observation, and enables cosmic surveys at the radio wavelengths, even with modest instruments with relatively small collection area. Using this enables us to collect information from the entire field-of-view of a full image to produce an estimate, and thus increases the signal-to-noise ratio. Thus, the visibility based method has the potential to detect large-scale, low-level radio emission like megahalos which may not be readily detectable in the image domain. We note that we could only measure the DGSE in the field for MACSJ0152 and we plan to systematically investigate this on a larger cluster sample in our future work. Further, ongoing and upcoming radio surveys with much higher sensitivity using interferometers like, VLA \citep{Nap83}, MeerKat \citep{Booth09}, ASKAP \citep{ASKAP10} and SKA are expected to produce huge amount of data and large samples of the clusters. Considering our estimator is designed to extract information while reducing the computational load \citep{samir14}, it may be a valuable alternative approach to study the ICM with the future data.

\section{Summary and Conclusions}
\label{sec:sec5}

The non-thermal, diffuse radio synchrotron emission detected from galaxy clusters indicates towards the existence of large-scale distribution of magnetic fields, and cosmic-ray particles in the ICM \citep{Fer96,Ferrari08,Fer12}. The reason why such large structures ($\sim {\rm Mpc}$ scales) of cosmic-ray particles should exist in the ICM is though unknown till date. Different models for explaining the presence of the ${\rm Mpc}$ scale radio halo include (1.) re-acceleration of the cosmic-ray electrons through turbulence in the ICM \citep{Brun01,Pet01,Pet08}, and (2.) the secondary cosmic-ray electrons produced from the cosmic-ray protons (through pion decay due to the interaction with ICM thermal protons) getting accelerated in shocks in the ICM \citep{Den80,Blasi99,Dolag00}. Further, the origin of the cosmic-ray electrons that fill the recently discovered large megahalo region and the phenomena responsible for filling such a large-scale structure are even more perplexing, and challenging to explain. Study of the galaxy cluster radio halo properties helps improving our understanding of the origin and evolution of cosmic-ray electron, magnetic field as well as astrophysical turbulence in the ICM.

In this work, we propose an alternative method to study the distribution of the magnetic fields and cosmic-ray electrons in the ICM \citep{Gov06,Raph06,Vik02,Dolag02} that can also potentially be a powerful method to study the turbulence. The angular power spectrum $\cl$ is an effective tool \citep{samir17a,Planck20} that characterizes the angular (or spatial) distribution of the specific intensity fluctuations in the ICM. The fluctuations in the specific intensity distribution of the non-thermal synchrotron radiation, in turn, depend on the underlying distribution of the magnetic fields, as well as the distribution and fraction of the cosmic-ray electrons present in the medium. Careful modelling of the diffuse synchrotron emission from the ICM, when compared with measured angular power spectrum from the radio observations may distinguish between underlying re-acceleration mechanism and the nature of turbulence in the ICM.

We have used the Tapered Gridded Estimator (TGE) on the visibility data to estimate the $\cl$. So far, TGE has been primarily used to study field-wide emissions of diffuse signals that fills the entire field-of-view. For emissions restricted within a small part of the PB, like the central halo of a galaxy cluster, we find that the underlying power spectrum may get suppressed due to the finite source size; the tapering due to TGE further modulate the amplitude of the power spectrum. We have estimated the this modulation factor in Section \ref{sec3b} to adapt the TGE for finite emissions. The finite size essentially modifies the normalisation factor in TGE, which here also depends on the size of the source, along with the effective window function of the TGE after tapering.

We have checked this in Section \ref{sec:simulation}, where we have used realistic simulations to validate the TGE for sources restricted within a small part of the PB. We have simulated the GMRT observation for Abell 2744, by considering the emission restricted within a small part of the PB as a multiplication of a Gaussian window with an underlying diffuse field having a power-law model power spectrum given by, $\cl^{M}\propto\ell^{-1}$. The width of the Gaussian window denote the observed size of Abell 2744. The visibilities for the sky emission, modulated by the GMRT primary beam pattern, have then been estimated for a baseline distribution that is identical to the actual observation considered here. The TGE is used on the simulated visibilities to estimate the power spectrum, where the analysis on these simulated data is identical to that on the real data. We have compared the $\cl$ from simulations with $\cl^{M}$ for validating our estimator. The recovered $\cl$ followed $\cl^{M}\propto\ell^{-1}$, but with suppressed amplitude (Figure \ref{Sim_1}). The suppression of $\cl^{M}$ depends on tapering parameter being used, and we observe higher suppression for lower tapering for the finite size of the source. This could be rectified using eq. (\ref{eq:c12}), the formalism derived in Section \ref{sec3b}. We could recover $\cl^{M}$ at all tapering within $\pm1\sigma$ error (cosmic variance) with high accuracy (Figures \ref{Sim_1} and \ref{Sim_2}), the fractional deviation being less than $20\%$ for $\ell\ge\ell^{\rm GC}_{\rm min}$ (Section \ref{sec:simulation}).

We report the measured $\cl$ for the synchrotron emission for the radio halo of Abell 2744. This is the first work of this kind where we have used the visibility correlations to study the turbulence in galaxy clusters and ICM, to the best of our knowledge. Previously turbulence in the ICM and magnetic field power spectrum have been studied in radio observations using RM, e.g. \cite{Gov06}, which are limited by sensitivity and reliability of the RM images. Using simulations, we could recover the estimated $\cl$ for Abell 2744 using small scale intensity fluctuations with a power-law power spectrum on top of an exponential radial surface brightness profile. The $\cl\propto\ell^{-3}$ for the underlying fluctuations possibly indicate turbulence in the halo region. Detailed comparison of the measured $\cl$ with sophisticated models of turbulence may explain this power-law. This, however, requires full 3-dimensional simulations of ICM turbulence from first principle and we plan to explore them in future using detailed, first principle numerical simulations. Following examples from literature \citep[e.g.][]{Tegmark96,Giar01,ali14,Saha19,Paul19}, we have also presented models and the best-fit parameters for the estimated $\cl$ in Table \ref{tab:tab_1} and Figure \ref{fig:fig3a}. We also plan to apply this method of estimating $\cl$ for a larger sample of galaxy clusters to check if the fluctuation power spectrum varies for sources with different merger histories and dynamical states. Our method of visibility-based power spectrum estimation is computationally very efficient, and may be valuable for studies with ngVLA and SKA which will generate massive amount of data in future.

\begin{acknowledgments}
We sincerely thank the anonymous reviewer for their thoughtful comments and constructive suggestions, which helped improve the quality of this manuscript. We thank Aritra Basu for his valuable comments, which greatly improved the manuscript. NR acknowledges support from the United States-India Educational Foundation through the Fulbright Program. SC would like to acknowledge SERB-Start-up Research Grant (SRG) and SERB-MATRICS for providing financial support.
\end{acknowledgments}

\section*{Data availability}
The original raw data used for this study are available in the GMRT archive. For any simulated data products described and used in this work, please contact SP and NR mentioning the requirements for the specific data and/or data products that are requested for.

\begin{contribution}

S. Pal contributed in execution, analysis; N.R. conceptualized the work; S.S., S. Paul and S.C. contributed in data analysis; T.S. contributed in simulation. All authors contributed to the interpretation of the results and the preparation of the manuscript.


\end{contribution}

%
\facilities{GMRT}




\appendix

\section{DGSE from Surrounding Fields}

\begin{figure}
\begin{center}
\includegraphics[scale=0.4]{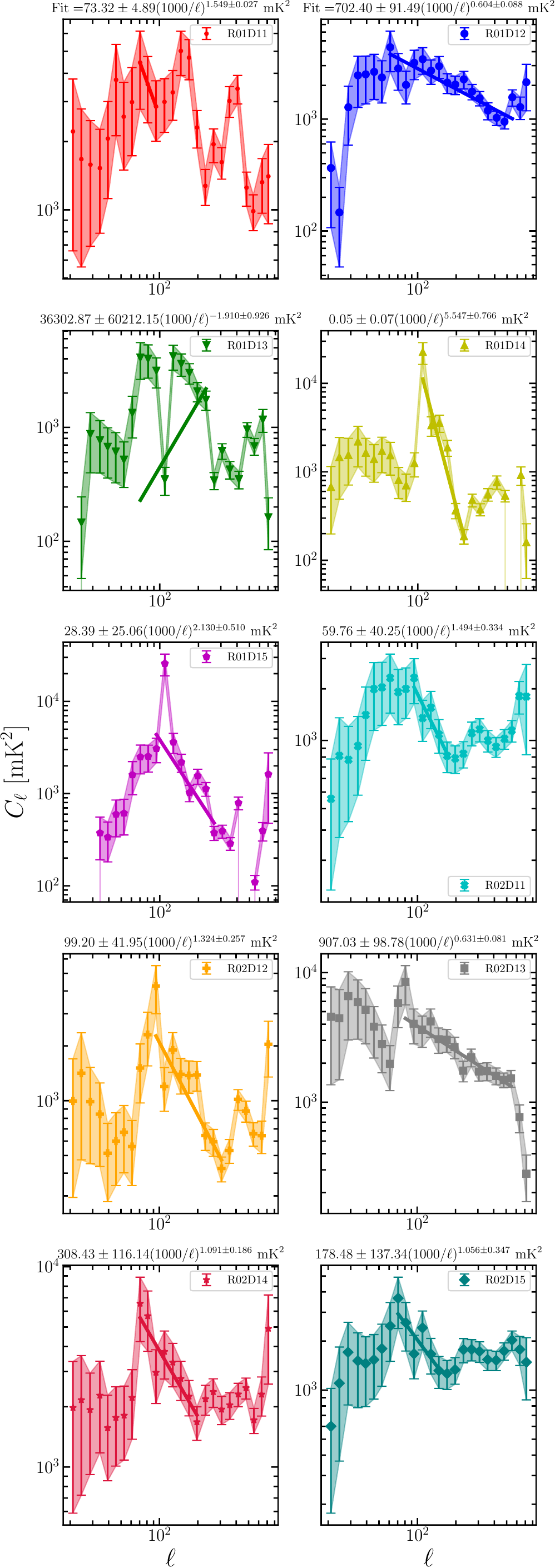}
\caption{DGSE fits for individual fields around Abell 2744, as shown in the left panel of Figure \ref{fig:Surroundingfields}. The $\cl$ values have been estimated in \citealt{samir20}. These best-fit parameters have been used for the DGSE predictions in Section \ref{DGSE} with the Parametric method.}
\label{fig:results_Abell}
\end{center}
\end{figure}

\begin{figure}
\begin{center}
\includegraphics[scale=0.4]{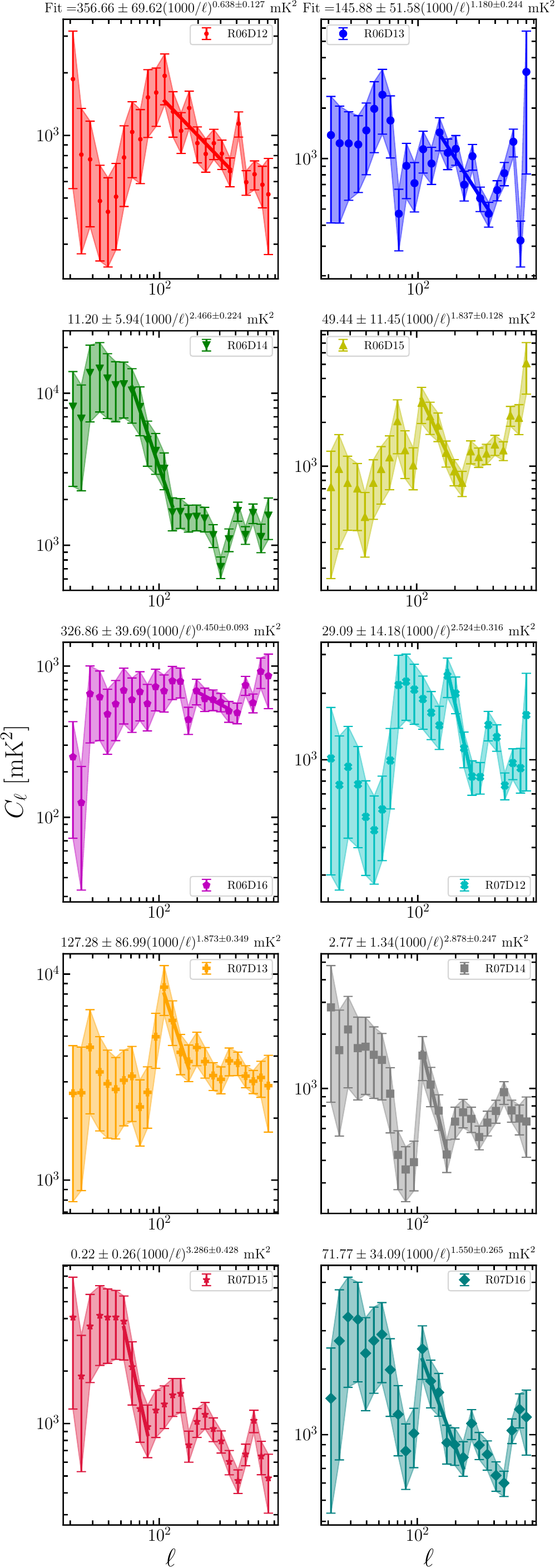}
\caption{DGSE fits for individual fields around MACSJ0152, as shown in the right panel of Figure \ref{fig:Surroundingfields}. The $\cl$ values have been estimated in \citealt{samir20}. These best-fit parameters have been used for the DGSE predictions in Section \ref{DGSE} with the Parametric method.}
\label{fig:results_MACSJ}
\end{center}
\end{figure}


\bibliography{myref}{}
\bibliographystyle{aasjournalv7}



\end{document}